\documentclass[preprint]{aastex}
\shorttitle{}
\shortauthors{}

\begin{document}

\title{Primordial Star Formation under Far-Ultraviolet Radiation}

\author{Kazuyuki Omukai}
\affil{Division of Theoretical Astrophysics, 
National Astronomical Observatory, Mitaka, Tokyo 181-8588, Japan}
\email{omukai@th.nao.ac.jp}

\begin{abstract}
Thermal and chemical evolution of primordial gas clouds irradiated with
far-ultraviolet (FUV; $h \nu < 13.6$ eV) radiation is investigated.
In clouds irradiated by intense FUV radiation, sufficient hydrogen
molecules to be important for cooling are never formed.  
However, even without molecular hydrogen, if the clouds are
massive enough, they start collapsing via atomic hydrogen line
cooling.
Such clouds continue to collapse almost isothermally owing to successive
cooling by H$^{-}$ free-bound emission up to the number density of
$10^{16}{\rm cm^{-3}}$. 
Inside the clouds, the Jeans mass eventually falls well below a solar
mass.
This indicates that hydrogen molecules are dispensable for low-mass
primordial star formation, provided fragmentation of the clouds
occurs at sufficiently high density. 
\end{abstract}


\section{Introduction}
Since the dawn of evolutionary cosmology, great efforts toward
understanding the formation of the first structures have been made by
many authors  
(e.g., Saslaw \& Zipoy 1967; Peebles \& Dicke 1968; Matsuda, Sato, \&
Takeda 1969; Doroshkevich \& Kolesnik 1976; Carlberg 1981; 
Palla, Salpeter, \& Stahler 1983; Kashlinsky \& Rees 1983).  
In particular, partly stimulated by observational advances and by the 
consequent detectability of the first stars by next-generation
facilities, this area is being vigorously investigated these days
(e.g., Haiman, Thoul, \& Loeb 1996; Ostriker \& Gnedin 1996; 
Tegmark et al. 1997; Yamada \& Nishi 1998; Bromm, Coppi, \& Larson 1999; 
Abel, Bryan, \& Norman 2000; Fuller \& Couchman 2000). 
Those studies have repeatedly emphasized the key role played by 
molecular hydrogen.

The role of H$_2$ in the cosmological context is twofold.
First, small primordial clouds whose virial temperature is 
$1000 \la T_{\rm vir} \la 8000{\rm K}$ can cool and start
collapsing via H$_2$ line cooling after virialization (e.g., Haiman
et al. 1996).   
Second, the Jeans mass inside a collapsing primordial cloud is
eventually reduced to much less than a solar mass as a result of
H$_2$ cooling (e.g., Palla et al. 1983). 

Regarding the former role, there is some dispute over the role of H$_2$ 
in the formation of second-generation stars, namely, what happens after the
first stars have formed by H$_2$ cooling.
Formed first stars or first quasars can be sources of UV radiation.
Whereas ionizing photons are trapped in an HII region around the source,  
far-ultraviolet (FUV; $h\nu <13.6$ eV) photons 
\footnote{Some authors use the term ``soft-UV'', instead of ``far-UV'', for
radiation whose energy is below the Lyman limit.} 
travel farther and form an outer H$_2$ photodissociation region (PDR).
Therefore, before the overlap of the HII regions, or in other words the 
reionization of the universe, the H$_2$ PDRs must cover the whole
intergalactic space. 
At that time, the universe is filled with FUV background radiation
(e.g., Haiman, Rees, \& Loeb 1997; Ciardi, Ferrara, \& Abel 2000). 

Hydrogen molecules in small objects can easily be photodissociated by
either internal or external UV sources.
Haiman et al. (1997) found that in the presence of UV
background radiation at the level needed to reionize the universe,
molecular hydrogen is photodissociated in small cosmological objects.   
Furthermore, Omukai \& Nishi (1999) pointed out that even a
single O star formed in a small cosmological object is able to
photodissociate the whole original cloud.
In addition, those clouds are easily blown out by a few supernovae because of
their shallow potential wells (e.g., Mac Low \& Ferrara 1999). 
These effects strongly regulate the subsequent star formation within
small pregalactic clouds. 

In contrast to this, the temperature of sufficiently massive clouds
reaches above $\sim$8000K, where atomic Ly$\alpha$ emission begins 
to be an effective cooling agent. 
Regardless of the H$_2$ photodissociation, such clouds can start
collapsing dynamically.
It is frequently assumed that, even in primordial clouds that cool by
the atomic cooling instead of the H$_2$ one, star formation occurs in
the same way as in those with the H$_2$ cooling. 
However, it is far from evident.
   
Recall another role of hydrogen molecules in primordial star formation.
The formation of protostars in primordial clouds with H$_2$ cooling has
been investigated by many authors (e.g., Matsuda et al. 1969; Yoneyama
1972; Carlberg 1981; Palla et al. 1983; Omukai \& Nishi 1998).
It is known that those clouds cool by H$_2$ rovibrational line
emission (for number density $n \la 10^{14} {\rm cm^{-3}}$), H$_2$
collision-induced emission (for $10^{14} \la n \la 10^{16} {\rm
cm^{-3}}$), and the H$_2$ dissociation (for $n \la 10^{20}{\rm cm^{-3}}$). 
Eventually, a very small protostar ($\sim 10^{-3} M_{\sun}$) 
forms at the center when $n \simeq 10^{22}{\rm cm^{-3}}$.  

On the other hand, we know little about the later evolution of clouds
that start collapsing via atomic cooling. 
It is still unclear under what condition H$_2$ eventually forms, or
whether it plays some role as a coolant in such clouds.
It is also not known how small a Jeans mass is reached through 
atomic cooling alone. 
If this value falls well beyond the mass of normal stars (for example if 
it corresponds to the size of super massive stars or very massive objects),
we must conclude that normal stars does not form only by atomic cooling.   
The aim of this paper is to answer these questions.

In this paper, we study the thermal and chemical evolution of primordial
clouds irradiated with FUV radiation.
Thereby, we demonstrate that the Jeans mass inside the clouds is
reduced to much less than a solar mass even in the absence of molecular
cooling.   
This implies that low-mass primordial star formation is possible even
without H$_2$ if fragmentation of the clouds occurs at sufficiently
high density.  

The outline of this paper is as follows.
In \S 2, the method of our calculations is described. 
In \S 3, results of our calculations are presented.
Finally, our work is summarized and its implications sketched
briefly in \S 4.

\section{Model}
We consider spherical clouds of hydrogen-helium gas irradiated with 
FUV radiation.
Specifically, we investigate the cases of two FUV spectra in
this paper: type {\it a}, the power-law radiation  
\begin{equation}
J_{\rm UV}(\nu)=J_{21} \times 10^{-21} (\nu/ \nu_{\rm th})^{-1} 
~{\rm (erg~s^{-1}~cm^{-2}~str^{-1}~Hz^{-1})}~~~(\nu < \nu_{\rm th}),
\end{equation}
and type {\it b}, the diluted thermal radiation of $10^{4}$ K 
\begin{equation}
J_{\rm UV}(\nu)=J_{21} \times 10^{-21} \frac{B(\nu; 10^{4} {\rm K})}{B(\nu_{\rm
th}; 10^{4} {\rm K})} 
~{\rm (erg~s^{-1}~cm^{-2}~str^{-1}~Hz^{-1})}~~~(\nu < \nu_{\rm th}), 
\end{equation}
where  $\nu_{\rm th}$ is the Lyman limit frequency.
In both cases, $J_{\rm UV}(\nu)=0$ beyond the Lyman limit.
We neglect the so-called sawtooth modulation due to the absorption of
Lyman series photons by the intergalactic matter (e.g., Haiman et al. 1997), 
for simplicity.  
This suffices for our simple analysis, which is valid only in 
order-of-magnitude estimates.   
We also include the cosmic microwave background radiation (CMBR), although it
has little influence on the matter at those low redshifts.
The spectra of FUV radiation are shown in Figure 1 for $J_{21}=1$ along
with the CMBR at $z=30$. 
Note that, for the same value of intensity at the Lyman limit, the number
of photons above the threshold of H$^{-}$ photodissociation is larger
for type {\it b} spectra than for type {\it a} spectra.  
Consequently, the photodissociation rate of H$^{-}$ is about 250
times larger for the $T=10^{4}$ K thermal type radiation (type {\it b}) than
for the power-law type (type {\it a}) with the same value of $J_{21}$.
Similarly, the photodissociation rate of H$_2$ is about 3 times larger
for the type {\it b} spectrum.

We take the maximum expansion of an overdensity as the initial condition
of the calculation.
Assuming the redshift of that epoch $z_{\rm m} \simeq 30$,
we take the number density of hydrogen nuclei $n=(\Omega_{\rm
b}/\Omega_{0})[\rho/(1+4y_{\rm He})m_{\rm H}]$, ionization degree
$y(e)$, and matter temperature $T_{\rm m}$ at the maximum expansion 
epoch as $n=8.9 \times 10^{-2} {\rm cm^{-3}}, T=39 {\rm K}, y(e)=2.0 \times
10^{-4}$, respectively.  
We neglect the initial molecular abundance and assume $y({\rm H_2})=0$.
Here $y_{\rm He}$ is the concentration of helium nuclei,
and $m_{\rm H}$ is the mass of a hydrogen nucleus.
Note that the concentration of He is defined by $y_{\rm He}=n_{\rm
He}/n$, where $n$ and $n_{\rm He}$ are the number density of hydrogen
nuclei and helium nuclei, respectively.
Simultaneously, we write for each atomic, molecular, or ionic species, 
$y(x)=n(x)/n$, where $n(x)$ is the number density of species $x$.
Note $y({\rm H_2})=1/2$ for fully molecular gas.
We set the primordial helium abundance at $y_{\rm He}=0.0807$, which
corresponds to the mass fraction $Y_{\rm p}=0.244$ (Izotov \& Thuan
1998).
The cosmological parameters are $\Omega_{\rm 0}=1, \Omega_{\rm b}=0.05$,
and $h=0.7$. 

Any effect due to rotation or magnetic fields are neglected for
simplicity.
Then, the actual collapse is expected to proceed like the
Penston-Larson similarity solution (Penston 1969; Larson 1969).
\footnote{Although the original Penston-Larson similarity solution is
limited to the isothermal collapse, Yahil (1983) extended this solution for
general polytropic equations of state.} 
According to this solution, the cloud consists of two parts, that is,
a central core region and an envelope.
The central core region has a flat density distribution, whereas in the
envelope, the density decreases outward as $\propto r^{-2}$.
The size of the central flat region is roughly given by the local Jeans
length $\lambda_{\rm J}=\pi c_{\rm s}/\sqrt{G \rho}$, where 
$c_{\rm s}$ and $\rho$ are the sound speed and total density in the core,
respectively. 
In this paper, we take the radius of the central region as $R_{\rm
c}=\lambda_{\rm J}/2$.
The collapse in the core proceeds approximately at the free-fall rate.
In this paper, we focus on the evolution in the central region.
In particular, we calculate the the temperature and chemical composition
of the collapsing core as a function of the central density.   

The cloud consists of two components, namely, baryon and dark matter.
We assume that the dynamics of the baryonic component is described by
the relation 
\begin{equation}
\frac{d \rho_{\rm b}}{dt}=\frac{\rho_{\rm b}}{t_{\rm ff}},
\end{equation}
where $\rho_{\rm b}$ is the baryonic density in the central region and
the free-fall time is
\begin{equation}
t_{\rm ff}\equiv \sqrt{\frac{3 \pi}{32 G \rho}}.
\end{equation}

The dynamics of the dark matter is described by the relation 
for the top-hat overdensity,
\begin{equation}
\label{eq:tophat}
\rho=\frac{9 \pi^{2}}{2} (\frac{1+z_{\rm m}}{1-{\rm cos} \theta})^{3}
\Omega_{0} \rho_{\rm c},
\end{equation}
up to the virialization,
where the maximum expansion redshift $z_{\rm m}=30$, and the parameter
$\theta$ is related to $z$ by the relation 
\begin{equation}
1+z=(1+z_{\rm m})(\frac{\theta - {\rm sin} \theta}{\pi})^{2/3}
\end{equation}
(e.g., Padmanabhan 1993).
We also use the usual age-redshift relation for the matter dominant
universe, $t=3.1 \times 10^{17}({\rm s}) h^{-1} \Omega_{0}^{-1/2}
(1+z)^{-3/2}$. 
After the virialization of dark matter, i.e., when the density reaches
to the virial density $\rho_{\rm DM, vir}=8 \rho_{\rm DM}(z_{\rm m})$,
we keep $\rho_{\rm DM}=\rho_{\rm DM, vir}$.

The thermal evolution is followed by solving the energy equation
\begin{equation}
\frac{d e}{dt}=-p \frac{d}{dt} (\frac{1}{\rho_{\rm b}})
- \frac{\Lambda_{\rm net}}{\rho_{\rm b}},
\end{equation}
where 
\begin{equation}
e=\frac{1}{\gamma_{\rm ad}-1} \frac{kT}{\mu m_{\rm H}}
\end{equation}
is the internal energy per unit mass of baryon, 
\begin{equation}
p=\frac{\rho_{\rm b} k T}{\mu m_{\rm H}}
\end{equation}
is the pressure for an ideal gas, $\gamma_{\rm ad}$ is the adiabatic exponent,
$T$ is the temperature, $\mu$ is the mean molecular weight, $m_{\rm H}$
is the mass of a hydrogen nucleus, and $\Lambda_{\rm net}$ is the net
cooling rate per unit volume. 
The net cooling rate $\Lambda_{\rm net}$ consists of contributions from
radiative cooling or heating by atomic hydrogen lines $\Lambda_{\rm H}$;
by rovibrational lines of molecular hydrogen, $\Lambda_{\rm H_2}$; by
continuum, $\Lambda_{\rm cont}$; and by the Compton coupling with the
radiation, $\Lambda_{\rm Compt}$
(the Compton cooling is unimportant in those low redshifts although it is
included in our calculation)
and from heating and cooling associated with chemical reactions, 
$\Lambda_{\rm chem}$:       
\begin{equation}
\Lambda_{\rm net}=\Lambda_{\rm H} +  \Lambda_{\rm H_2} + \Lambda_{\rm cont}
+ \Lambda_{\rm Compt}+ \Lambda_{\rm chem}.
\end{equation}
The continuum processes included are listed in Table 1.
The details of these processes are described in Appendix B.
Since we are focusing on the evolution of the central region, whose radius
$R_{\rm c}=\lambda_{J}/2$, then we evaluate the optical depth
$\tau_{\nu}$ by 
\begin{equation}
\tau_{\nu}=\kappa_{\nu} R_{\rm c}= \kappa_{\nu}(\frac{\lambda_{\rm
J}}{2}).
\end{equation}

Time-dependent nonequilibrium chemical reactions are solved for the
following nine species:
H, H$_{2}$, $e$, H$^{+}$, H$_{2}^{+}$, H$^{-}$, He, He$^{+}$, and
He$^{++}$. 
Chemical reactions included are listed in Table 2.

\section{Results}
In this section, we present our numerical results.
Figures 2 and 3 display the temperature evolution for collapsing primordial
clouds irradiated with (1) the power-law type and (2) diluted black body of
$10^{4}$ K type FUV radiation, respectively. 
As mentioned in \S 2, the photodissociation rate coefficients of
both H$^{-}$ and H$_2$ are larger for type {\it b} spectrum than
for the type {\it a} with the same value of $J_{21}$.
Consequently, FUV radiation of type {\it b} has the larger influence
on the evolution of the clouds than type {\it a} with the same value of
$J_{21}$.  
We discuss effects of different spectral types and scaling relations
between them in Appendix A.  
Except for those scalings of $J_{21}$, the evolutionary
trajectories change in similar ways with an increase of the UV intensity 
in both cases. 
Hence, we describe only the evolutionary features of the clouds
irradiated with type {\it a} radiation in detail here.

The evolutionary trajectories for the clouds with $J_{21} \geq 10^{5}$ are
identical to each other.
As is obvious from Figure 2, the thermal evolution at high densities
(say, $n \ga 10^{7} {\rm cm^{-3}}$) is completely different between clouds with
$J_{21} \leq 10^{4}$ and those with $J_{21} \geq 10^{5}$.
This is because the clouds irradiated by the FUV radiation with $J_{21}
\geq 10^{5}$ cannot form sufficient ${\rm H_2}$, so they cool only
by atomic cooling.
The temperature of such clouds is higher than that of clouds
with $J_{\rm 21} \leq 10^{4}$, which can form enough ${\rm H_2}$ to
cool eventually.
Toward higher densities, these two groups of trajectories in Figure 1
converge respectively to two different limiting tracks.
We call the higher temperature one the ``atomic cooling track'',
and the other the ``molecular cooling track''.
This nomenclature comes from the fact that clouds evolving along the
former track remain atomic, while those following the latter eventually 
become fully molecular. 
We should note that the Jeans masses inside the clouds ultimately fall below 
$0.1 M_{\sun}$ for both tracks. 

Figure 4 shows the fractional abundances of hydrogen molecules (Fig.4{\it a})
 and electrons (Fig.4{\it b}) for the clouds irradiated with type {\it a} 
radiation. 
The separate heating and cooling rates per unit mass
for those clouds are illustrated in the panels of Figure 5 for
$J_{21}=0$(Fig.5{\it a}), $10^{4}$(Fig.5{\it b}) and $10^{5}$(Fig.5{\it c}).

We review their evolutionary features starting from the case of $J_{21}=0$.
In this case, the collapse proceeds along the well-known molecular cooling
track (e.g., Palla et al. 1983).
At the beginning, the temperature rises adiabatically owing to the
compression and the initial lack of coolant. 
If the total mass of the cloud is less than the maximum Jeans
mass attained at the end of this adiabatic phase, the contraction stops
at the time when the cloud mass becomes equal to the Jeans mass. 
We assume here that the total mass of the cloud is larger than the 
maximum Jeans mass.
At low densities (i.e., $n \la 10^{8}{\rm cm^{-3}}$), H$_2$ is formed mainly
through the H$^{-}$ channel:
\begin{equation}
{\rm H + {\it e} \rightarrow H^{-} + \gamma}~~~~{\rm (reaction~32)},
\end{equation} 
and then
\begin{equation}
{\rm H + H^{-} \rightarrow H_2 + {\it e}}~~~~{\rm (reaction~9)}.
\end{equation} 
Increased temperature and density causes an increase of both the H$_2$
formation rate and the H$_2$ cooling function.   
Consequently, a sufficient amount of H$_2$ to cool within a free-fall
time is formed, and the temperature then drops.
A detailed discussion of when the efficient H$_2$ cooling begins is 
presented in Appendix A.  
Owing to the H$_2$ cooling, the temperature is kept as low as
a few hundred K in the cloud (see Fig.2 and Fig.5{\it a}). 
When the density reaches about $10^{8}{\rm cm^{-3}}$, the
three-body reaction 
\begin{equation}
{\rm 3 H \rightarrow H_2 + H}~~~({\rm reaction~16})
\end{equation}
becomes efficient (Palla et al. 1983).
In consequence of this highly productive H$_2$ formation, hydrogen
becomes fully molecular (see Fig.4 {\it a}).
At the same time, the cloud becomes optically thick to a few H$_2$
lines at $n \simeq 10^{11} {\rm cm^{-3}}$.
However, the cooling remains efficient enough to induce dynamical
collapse (i.e., The ratio of specific heat $\Gamma \equiv {\rm
dlog}p/{\rm dlog}\rho < 4/3$) because there are still enough optically
thin lines.  
When the central number density reaches about $10^{14} {\rm cm^{-3}}$,
H$_2$ collision-induced emission begins to dominate the cooling (see 
Fig.5{\it a}).  
The cloud becomes optically thick to this continuum at $\sim 10^{16}
{\rm cm^{-3}}$. 
Thereafter, the radiative cooling rate drops rapidly.
Simultaneously, H$_2$ dissociation begins, because the temperature is
already near the value needed for dissociation. 
The H$_2$ dissociation prevents the temperature from rising rapidly until the
number density reaches about $10^{20} {\rm cm^{-3}}$.
The minimum Jeans mass of $1.5 \times 10^{-3} M_{\sun}$ is reached
shortly after that. 
When most of the hydrogen molecules are dissociated, the collapse
becomes approximately adiabatic.
According to Omukai \& Nishi (1998), after a little further contraction, 
a small hydrostatic core (about a few $10^{-3} M_{\sun}$), or
protostar, forms at $n \simeq 10^{22} {\rm cm^{-3}}, T \simeq 3 \times
10^{4}$K. 

As seen in Figure 2, if the intensity of FUV radiation increases, the
density and temperature at the time when efficient H$_2$ cooling begins
also increases (see Appendix A for further discussion).   
After H$_2$ cooling becomes effective, the trajectories of those clouds 
soon converge to the evolutionary track for $J_{\rm UV}(\nu)=0$.

An interesting example is when the intensity is as high as $J_{21}
\simeq 10^{4}$. 
In this case, the temperature rises almost adiabatically up to about
8000K without enough H$_2$ formation.
At that point, Ly$\alpha$ line emission of atomic hydrogen begins to
work as an efficient cooling agent (see Fig.5{\it b}).
After the cloud collapses isothermally at $T\simeq 8000$K over an order
of magnitude in density owing to this coolant, sufficient 
hydrogen molecules form as a result of the increased density.
Thereafter, the trajectory converges rapidly to the molecular cooling
track in the same way as those with weaker radiation (see Fig.2).

When the intensity of FUV radiation is increased more, 
i.e., $J_{21} \ga 10^{5}$, the thermal evolution becomes
completely different from those with lower intensities (see Fig.2).
The adiabatic rise of temperature continues until Ly$\alpha$
cooling becomes effective at about 8000K.  
\footnote{
The initial slight rise of ionization degree seen in Figure 4{\it b} is
caused by photoionization from the first excited level, which is populated 
by more atoms than the local thermodynamic equilibrium (LTE) value as a
result of absorption of irradiated Ly$\alpha$ photons. 
This effect has little influence on the later evolution.
}
While the cloud collapses nearly isothermally at about 8000K owing to
Ly$\alpha$ line cooling, the central density reaches the critical
density 
\footnote{The critical density is the density above which the
rovibrational levels of H$_2$ are populated, according to the LTE law.} 
$n_{\rm cr} \simeq 10^{4} {\rm cm^{-3}}$ of hydrogen molecules
before a amount of H$_2$ sufficient to cool within a free-fall time 
forms.
At higher densities, $n \ga n_{\rm cr}$, enough H$_2$ never forms, for
the following two reasons.
First, the H$_2$ fraction decreases;
the collisional dissociation rate coefficient increases because, 
at $n \ga n_{\rm cr}$, high vibrational levels, at which dissociation  
occurs easily, are populated by more hydrogen molecules than at lower
density.
Second, the amount of H$_2$ needed for efficient cooling increases;
note that the cooling rate per unit volume $\Lambda_{\rm H_2} \propto
n^{2}$ for $n \la n_{\rm cr}$, while $\Lambda_{\rm H_2} \propto n$ for
$n \ga n_{\rm cr}$. 
(see Appendix A for further discussion).
Consequently, the cloud never joins to the molecular cooling track.
Instead, it continues to cool by Ly$\alpha$ line emission until the density
reaches about $10^{6} {\rm cm^{-3}}$, where collisional de-excitation
from the $2p$ state begins to dominate the Ly$\alpha$ emission as a
result of the small escape probability (i.e., $\beta_{{\rm esc},21}
A_{21} \simeq C_{21}$). 
The subsequent major cooling process is two-photon emission of atomic
hydrogen from the $2s$ state.
This works effectively in the density range $10^{6}{\rm cm^{-3}} \la n
\la 10^{7}{\rm cm^{-3}}$.
Next, the cloud cools by the free-bound emission of ${\rm H^{-}}$ 
over nearly ten orders of magnitude in density, until the number density
reaches $10^{16} {\rm cm^{-3}}$ (see Fig.5 {\it c}).
The mechanism of ${\rm H^{-}}$ free-bound emission cooling is as
follows.
As a first step, radiative association of H and $e$;
\begin{equation}
 {\rm H + {\it e} \rightarrow H^{-} + \gamma}~~~({\rm reaction~32})
\end{equation}
occurs and a photon is emitted.
This ${\rm H^{-}}$ is used to form H$_2$ (reaction 9), which will be
collisionally dissociated by reaction 15.
As a result, there is net cooing by the photon emitted in reaction 32.
\footnote{Formed ${\rm H^{-}}$ is photodissociated at low densities, that is, 
$n< k_{31}/k_{9}= 0.1 \alpha J_{21}$, where $\alpha$ is a constant
defined in Appendix A, and is 8 for the type {\it a} ($2 \times
10^{3}$ for type {\it b}, respectively) spectrum.    
In this case, there is no net cooling.}
In the course of the dynamical collapse induced by H$^{-}$ free-bound emission,
the temperature falls gradually from 7000K (at $n \simeq 10^{6} {\rm
cm^{-3}}$) to 3000K (at $n \simeq 10^{16} {\rm cm^{-3}}$; see Fig. 2).
At the central density, $\sim 10^{16} {\rm cm^{-3}}$, the cloud becomes
optically thick to both H$^{-}$ bound-free absorption and Rayleigh scattering
of atomic hydrogen.  
At the number density $\simeq 10^{17} {\rm cm^{-3}}$, the H ionization
begins (see Fig. 4 b).
As a result, the temperature rise slows down up to $n \simeq 10^{19} {\rm
cm^{-3}}$. 
At $n=10^{20} {\rm cm^{-3}}$, where 60 \% of the gas is ionized, the
minimum Jeans mass $M_{\rm J, min}=0.03 M_{\sun}$ is reached.  
Thereafter, the temperature continues to rise almost adiabatically.
We expect that a protostar on the order of $M_{\rm J, min}$ is formed
after a little further adiabatic contraction.
Although the minimum Jeans mass of the atomic cooling clouds is 20
times larger than that of the H$_2$ cooling clouds, it is still much less
than the mass of usual stars. 

\section{Summary and Discussion}
We have investigated the thermal and chemical evolution of primordial
clouds irradiated with FUV radiation.
When intensity of the irradiating FUV radiation exceeds some critical
value [$J_{21} \simeq 10^{5}$ for $J_{\rm UV}(\nu) \propto \nu^{-1}~~( h \nu <
13.6 {\rm eV})$], 
sufficient molecular hydrogen to be important for cooling is never
formed in those clouds because of the photodissociation and blocking of
the formation channel.  
Nonetheless, sufficiently massive clouds can start dynamical
collapse by atomic line cooling even without H$_2$.
Those clouds continue to collapse almost isothermally at several
thousand K along the ``atomic cooling track'' in the density-temperature
plane.
That dynamical collapse is induced by successive cooling by
Ly$\alpha$ emission, two-photon emission, and H$^{-}$ free-bound
emission.
The minimum Jeans mass eventually reduces to about $0.03 M_{\sun}$ for
those clouds. 
On the other hand, clouds irradiated with less intense FUV radiation
collapse dynamically by the H$_2$ cooling.  
The minimum Jeans mass of such clouds is about $1.5
\times 10^{-3} M_{\sun}$.

Although in this paper we limited our application to the protostellar
collapse under an FUV background, we expect that the primordial protostellar
collapse without H$_2$ also occurs in other situations.
Recently, Susa \& Kitayama (2000) investigated the collapse of
primordial clouds irradiated with ionizing UV background
$J_{\rm UV}(\nu)=J_{21} \times 10^{-21}(\nu/\nu_{\rm th})^{-1}~{\rm (erg~s^{-1}
cm^{-2} str^{-1} Hz^{-1})}$.  
They pointed out that H$_2$ formation is suppressed for $J_{21}>3.6
\times 10^{4}$ owing to blocking of the H$_2$ formation channel by H$^{-}$
photodissociation.
Although they did not study the later evolution, we guess that the
clouds continue collapsing along our atomic cooling track. 
Carlberg (1981) also considered the thermal evolution of collapsing
primordial clouds without external radiation, but he did not include
the three-body process of ${\rm H_2}$ formation.
For this reason, in his calculation, molecular hydrogen dissociates at
$n \simeq 10^{11} {\rm cm}^{-3}$.
After that, the evolutionary path of his cloud follows approximately our
atomic cooling track (see Fig. 2 of Carlberg 1981). 
This fact support our speculation that the clouds that do not cool by
H$_2$ collapse along the atomic cooling track.

Next, we discuss the feasibility of the collapse along the atomic cooling
track in the cosmological reionization.
The UV intensity needed to reionize the universe at $z_{\rm reion}$,
i.e., 1 ionizing photon per a hydrogen nucleus, is   
\begin{equation}
J_{\rm reion}=\frac{hc \Omega_{\rm b} \rho_{\rm cr} (1+z_{\rm
reion})^{3}}{4 \pi (1+4 y_{\rm He}) m_{\rm H}}=3 \times 10^{-21}
(\frac{1+z_{\rm reion}}{10})^{3} 
~{\rm (erg~s^{-1}~cm^{-2}~str^{-1}~Hz^{-1})}
\end{equation}
at the Lyman limit for $J(\nu) \propto \nu^{-1}$ and the assumed
cosmological parameters.
Thus, the condition that the clouds collapse along the atomic
cooling track, $J_{21} \ga 10^{5}$ for $J_{\rm UV}(\nu) \propto
\nu^{-1}$, seems to be rarely met in practice, except in the immediate
vicinity of UV sources.  
 
Finally, we discuss the mass of formed stars.
Although the minimum Jeans mass is less than 0.1 $M_{\sun}$, this
corresponds not to the final mass of formed stars, but rather to the initial
mass of the protostar, which has no direct relation with the former. 
When a small hydrostatic core, namely protostar, forms at the center of
a collapsing protostellar cloud, an enormous amount of gravitationally 
unstable gas still surrounds the protostar.    
Thereafter, the protostar grows in mass by accreting the ambient matter.
Consequently, the final mass of stars is determined by the subsequent
accretion onto the protostar.
Stahler, Shu, \& Taam (1980) argued that a rough estimate of the
protostellar mass accretion rate ${\dot M}$ can be obtained from the
relation ${\dot M} \sim c_{\rm s}^{3}/G$,
where $c_{\rm s}$ is the isothermal sound speed in the protostellar
cloud.
From this relation, the mass accretion rate is higher
for protostars formed by atomic cooling than for those formed by molecular
cooling because of the higher temperature of the protostellar cloud.
Suppose that the accretion is halted by some stellar activity, e.g., 
radiation force, bipolar flow, etc.
In this case, the higher accretion rate probably results in the later
halting and higher mass of formed stars (e.g., Wolfire \& Cassinelli
1987).

In contrast, from the viewpoint of fragmentation, the masses of
primordial stars formed by atomic cooling alone could be smaller than
those formed by molecular cooling.
The fragmentation is possible in principle as long as the temperature 
decreases toward higher density, while it does not occur after the
isothermality of the cloud breaks down and the temperature begins to
rise (Inutsuka \& Miyama 1997; Masunaga \& Inutsuka 1999). 
From Figures 2 and 3, the final fragmentation occurs at about $n \simeq
10^{16} {\rm cm^{-3}}$ for clouds collapsing along the atomic cooling track.
On the other hand, it seems difficult for molecular cooling clouds to
fragment at $n \ga 10^{3} {\rm cm^{-3}}$.
The masses of fragments, which is about the Jeans mass at that time, are
about $0.1 M_{\sun}$ for the atomic-cooling clouds, and $10^{4} M_{\sun}$
for the molecular-cooling clouds.
If this is the case and the whole material contained in a fragment is
converted into a single star, very massive stars form from 
a molecular-cooling cloud, and red dwarfs form as a result of atomic cooling.  
However, it should be noted that the fragmentation and thermal evolution
depend strongly on each other and also on the geometry of the clouds. 
In fact, Uehara et al. (1996) concluded that the minimum mass of
fragments is on the order of a solar mass, which is essentially the
Chandrasekhar mass, by studying thermal evolution of filamentary
primordial clouds that collapse by molecular cooling. 
Later, improved simulations by Nakamura \& Umemura (1998) confirmed
Uehara et al.(1996)'s result. 
The discrepancy between our mass of fragments ($\sim 10^{4} M_{\sun}$)
and theirs ($\sim 1 M_{\sun}$) results from the difference of the
assumed geometry of the clouds; 
we have assumed spherical clouds in studying the thermal evolution,
while their clouds are filamentary. 
For filamentary clouds, the gravitational contraction is slower and
compressional heating rate is lower than for spherical ones.
Hence, for the filamentary cloud the isothermality breaks down at higher 
density, where the Jeans mass is smaller. 
To fully address the complexity arising from the cloud geometry,
two-dimensional or three-dimensional calculations are needed.  
Furthermore, there are other uncertainties relating to the evolution of
the fragments, for example, mass accretion onto the fragments,
merger between them, feedback from neighboring formed stars, etc. 
(e.g., Bromm et al. 1999).
It seems premature to draw conclusions here about these issues.

\acknowledgements
I would like to thank Shu-ichiro Inutsuka, Ryoichi Nishi,
Humitaka Sato, Toru Tsuribe for crucial suggestions and discussions, 
and Phil Stancil for providing the data of the H$_2^{+}$
photodissociation cross section. 
I also acknowledge the referee for improving the manuscript.
This work is supported in part by Research Fellowships of the Japan
Society for the Promotion of Science for Young Scientists, grant 4287. 
\newpage
\appendix
\section{Conditions for H$_2$ Cooling}
In this section, we discuss key processes determining the amount of 
the molecular hydrogen forming under FUV radiation.
In particular, we delineate the condition for sufficient H$_2$ for
cooling to form.  

The H$_2$ fraction $y({\rm H_2})$ necessary to cool within a
free-fall time is given by 
\begin{equation}
\label{eq:y_need}
y_{\rm cool}({\rm H_2})=\frac{(3/2) k_{\rm B} T}{n L_{\rm H_2}
t_{\rm ff}},
\end{equation}
where $L_{\rm H_2} ({\rm erg~s^{-1}~cm^{3}})=\Lambda_{\rm H_2}/n({\rm
H})n({\rm H_2})$ is the H$_2$ cooling function and we assume $n({\rm
H})\simeq n$.  
When the actual H$_2$ fraction $y({\rm H_2})$ reaches $y_{\rm cool}({\rm
H_2})$, the temperature drops abruptly as seen in Figure 2 and 3.

Next, we discuss how much H$_2$ forms during the collapse.
The H$_2$ abundance is bounded by the following three values;

(1) the amount of H$_2$ that is able to form within a free-fall time,
$y_{\rm form}({\rm H_2})$, 

(2) the chemical equilibrium value between H$_2$ formation and 
photodissociation, $y_{\rm pd}({\rm H_2})$,

(3) the chemical equilibrium value between H$_2$ formation and 
collisional dissociation, $y_{\rm cd}({\rm H_2})$.

In practice, the smallest among these three values is realized;
\begin{equation}
\label{eq:yH2} 
y({\rm H_2})={\rm min}[y_{\rm form}({\rm H_2}),y_{\rm pd}({\rm
H_2}),y_{\rm cd}({\rm H_2})].
\end{equation}
In the following, we examine the values of $y_{\rm form}({\rm
H_2}),y_{\rm pd}({\rm H_2})$, and $y_{\rm cd}({\rm H_2})$ in this order. 

First, we find the formation time limited value $y_{\rm form}({\rm
H_2})$.
H$_2$ formation occurs mainly through the H$^{-}$ channel: 
\begin{equation}
{\rm H + {\it e} \rightarrow H^{-} + \gamma}~~~~{\rm (Reaction~32)},
\end{equation} 
followed by
\begin{equation}
{\rm H + H^{-} \rightarrow H_2 + {\it e}}~~~~{\rm (Reaction~9)}.
\end{equation} 
In the above chain, not all of H$^{-}$ formed are used in H$_2$ formation.
Instead, some of H$^{-}$ are photodissociated via Reaction 31;
\begin{equation}
{\rm H^{-} + \gamma \rightarrow H + {\it e}}~~~~{\rm (Reaction~31)}.
\end{equation} 
The photodissociation rate coefficient of H$^{-}$, $k_{31}$, depends on the
spectrum of radiation.
Here, we introduce a parameter $\alpha$, defined by
\begin{equation}
k_{31}=\alpha \kappa_{31} J_{21},
\end{equation}
where $\kappa_{31}=1 \times 10^{-10}$ is the value of $k_{31}$ under
$J_{\rm UV}(\nu)=10^{-21}={\rm const.}$ (for $\nu < \nu_{\rm th}$).
The parameter $\alpha$ characterizing the radiation spectrum above H$^{-}$
photodissociation threshold (0.755 eV) is 8
for the power-law type spectrum (type {\it a} in the text), $2 \times 10^{3}$
for the diluted thermal type spectrum (type {\it b}).  
The rates of the competing reactions 9 and 31 determine the branching
ratio of H$^{-}$ to be used in H$_2$ formation, which is
$k_{9}n/(k_{9}n+k_{31})$. 
Then, by defining the effective H$_2$ formation rate coefficient
\begin{equation}
k_{\rm form}=k_{32} \frac{k_{9}n}{k_{9}n+k_{31}},
\end{equation}
the H$_2$ formation rate can be written as $k_{\rm form} n(e) n({\rm
H})$.
Using $k_{\rm form}$ above, we obtain the H$_2$ fraction that can form
within a free-fall time 
\begin{equation}
\label{eq:y_f}
y_{\rm form}({\rm H_2})=k_{\rm form} y(e) n t_{\rm ff}.
\end{equation}

Second, we find the photodissociation limited value $y_{\rm pd}({\rm
H_2})$.
The photodissociation occurs via 
\begin{equation}
{\rm H_2 + \gamma \rightarrow H_2^{\ast} \rightarrow 2 H}~~~{\rm
(Reaction~35)},
\end{equation}
whose reaction
coefficient is
\begin{equation}
k_{\rm pd} \equiv k_{35}= 10^{9} J(h\nu=12.4 {\rm eV}) f_{\rm sh}
= 10^{-12} \beta J_{21} f_{\rm sh}. 
\end{equation}
Here, we have introduced a parameter $\beta$ that represents the relative
strength of radiation intensity at the average LW
band frequency (12.4 eV) to that at the Lyman limit (13.6 eV).
The shielding factor $f_{\rm sh}$ is given by
(Draine \& Bertoldi 1996)
\begin{equation}
\label{eq:fsh}
f_{\rm sh}={\rm min} \left[ 1,(N_{\rm H_2}/10^{14}{\rm cm^{-2}})^{-3/4}
\right].
\end{equation}
The equilibrium value between the H$_2$ formation and photodissociation
is
\begin{equation}
\label{eq:ypd}
y_{\rm pd}({\rm H_2})=\frac{k_{\rm form}}{k_{\rm pd}} y(e) n.
\end{equation}
If $N_{\rm H_2} >10^{14} {\rm cm^{-2}}$, the coefficient 
$k_{\rm pd}$, and then the right hand side of equation (\ref{eq:ypd}),
involves $y({\rm H_2})$ itself. 
Denoting $k_{\rm pd}=a_{\rm pd} y({\rm H_2})^{-3/4}$, where $a_{\rm pd}
\equiv 7 \times 10^{-18} T_{3}^{-3/8} n^{-3/8} \beta J_{21}$,  
we obtain
\begin{equation}
\label{eq:y_pd_sh}
y_{\rm pd}({\rm H_2})=\left( \frac{k_{\rm form}}{a_{\rm pd}} y(e) n
\right)^{4}~~~~~(N_{\rm H_2} >10^{14} {\rm cm^{-2}}). 
\end{equation}
Combining expressions for cases with and without H$_2$ self-shielding,
the photodissociation limited value of H$_2$ is
\begin{equation}
\label{eq:y_pd}
y_{\rm pd}({\rm H_2})={\rm max} \left[ \frac{k_{\rm form}}{10^{-12}
\beta J_{21}} y(e) n, \left( \frac{k_{\rm form}}{a_{\rm pd}} y(e) n
\right)^{4} \right]. 
\end{equation}
Note that the radiation intensity enters into the expression for $y_{\rm
pd}({\rm H_2})$ only in the form of $\alpha \beta J_{21}^{2}$
in the low density range $n<k_{31}/k_{9} \simeq 0.1 \alpha
J_{21} {\rm cm^{-3}}$. 

Third, the collisional dissociation limited value $y_{\rm cd}({\rm
H_2})$ can be obtained as follows.
The dominant collisional dissociation process is collision with the atomic
hydrogen: 
\begin{equation}
{\rm H_2 + H \rightarrow 3 H}~~~({\rm Reaction~15}).
\end{equation}
Then, denoting $k_{\rm cd} \equiv k_{15}$,
\begin{equation}
\label{eq:y_cd}
y_{\rm cd}({\rm H_2})=\frac{k_{\rm form}}{k_{\rm cd}}y(e).
\end{equation}
This value $y_{\rm cd}({\rm H_{2}})$ becomes smaller when the density exceeds
the critical density $n_{\rm cr} \simeq 10^{4} {\rm cm^{-3}}$ owing to 
the collisional dissociation from high vibrational states.

Now, we examine when sufficient H$_2$ forms under FUV radiation in
the case that the temperature rises adiabatically from the initial value.
Figure 6 shows the H$_2$ fraction estimated by Equation
(\ref{eq:yH2}) in this case.
Here, the temperature is given by $T=T_{0}(n/n_{0})^{2/3}$,
where the initial number density and temperature are $n_{0}=8.9 \times
10^{-2} {\rm cm^{-3}}, T_{0}=39 {\rm K}$, and the ionization degree
$y(e)$ is fixed to $10^{-4}$.
\footnote{Here and through out this section, we assume the ionization
degree is constant, namely, the recombination and/or ionization time
scale is longer than the evolutionary timescale of the system (i.e.,
free-fall time)
Although this is somewhat oversimplification, it suffices for clarifying 
key processes determining the H$_2$ fraction. }
In Figure 6, each curve for $y({\rm H_2})$ with fixed FUV intensity
consists of four segments.
From the lowest to highest temperature in this Figure, those segments
correspond to $y_{\rm pd}({\rm H_2})$ without  
self-shielding (gradually increasing portion), $y_{\rm pd}({\rm H_2})$
with self-shielding (rapidly increasing portion), $y_{\rm form}({\rm
H_2})$ (another gradually increasing portion), and  $y_{\rm cd}({\rm H_2})$ 
 (decreasing portion).
As known from Figure 6, except for lowest intensities (in such a case,
the evolution of the cloud is virtually unaffected by the presence of
FUV radiation), the photodissociation limited
value $y_{\rm pd}({\rm H_2})$ determines when the sufficient H$_2$ forms.
The condition that molecular cooling becomes effective before
atomic cooling does is that $y_{\rm pd}({\rm H_2})> y_{\rm cool}({\rm
H_2})$ at $T=8000$ K, which leads to 
\begin{equation}
J_{21}<4 \times 10^{3} (\frac{x_{4}}{\alpha \beta})^{1/2},
\end{equation}
where $x_{4} \equiv y(e)/10^{-4}$ and we have used 
$k_{\rm form} \simeq k_{32}k_{9}n/k_{31}$, and
$k_{32}=3.6 \times
10^{-15}, k_{9}=8.7 \times 10^{-10},y_{\rm cool}({\rm H_2})=10^{-6}$,
and $n=2.6 \times 10^{2} {\rm cm^{-3}}$ at 8000K.
Also, since FUV intensity enters into $y_{\rm pd}({\rm H_2})$ in the
form of $ \alpha \beta J_{21}^{2}$, type {\it b} spectrum of $J_{21}^{({\rm
b})}$ has the same effect as type {\it a} spectrum of $\sqrt{\alpha \beta
({\rm b}) /\alpha \beta ({\rm a})} \simeq 27$ times
$J_{21}^{({\rm b})}$ on the early evolution of the clouds. 

As discusses in the text, the cloud irradiated by strong FUV radiation 
never forms sufficient H$_2$ for cooling if the intensity exceeds a 
certain critical value.   
Next, we discuss the physical reason where this value
comes from.
Here, we examine how much H$_2$ forms while the cloud collapses
isothermally owing to Ly$\alpha$ cooling.
The values of H$_2$ concentration $y({\rm H_2})$ estimated by equation
(\ref{eq:yH2}) for $T=8000$K and $y(e)=10^{-4}$ are shown in Figure 7.
\footnote{Once the actual H$_2$ fraction exceeds $y_{\rm cool}({\rm H_2})$, 
the temperature of the cloud drops from $\sim 8000$K as a result of the
H$_2$ cooling.
After that, the H$_2$ fraction shown in the Figure does not represent
the actual value.}
For a given value of FUV intensity, $y({\rm H_2})$ is limited by $y_{\rm
pd}({\rm H_2})$ (an almost vertical line) in the lower density range and
by $y_{\rm cd}({\rm H_2})$ (a smooth curve) in the higher
density range. 
The formation time limited value $y_{\rm form}({\rm
H_2})$ does not appear in this Figure.
On the other hand, $y_{\rm cool}({\rm H_2})$ decreases in the low
density regime, reaches the minimum at $n \simeq n_{\rm cr} \simeq
10^{4} {\rm cm^{-3}}$, and increase in higher densities.	 
As seen in Figure 7, the H$_2$ fraction $y({\rm H_2})$ has no chance to
reach $y_{\rm cool}({\rm H_2})$ in $n \ga n_{\rm cr} \simeq 10^{4} {\rm
cm^{-3}}$.
This is because the H$_2$ fraction $y({\rm H_2})$ is limited by $y_{\rm
cd}({\rm H_2})$, which becomes small in those densities owing to the
enhanced collisional dissociation from high vibrational levels. 
Therefore, in order for the H$_2$ cooling to become efficient, 
 $y({\rm H_2})$ must reach $y_{\rm cool}({\rm H_2})$ at $n \la n_{\rm
cr}$; both $y_{\rm cd}({\rm H_2})$ and $y_{\rm pd}({\rm H_2})$ must
exceed $y_{\rm cool}({\rm H_2})$ at $n \la n_{\rm cr} \simeq 10^{4}{\rm
cm^{-3}}$.
As the condition on $y_{\rm cd}({\rm H_2})$, 
we take $\gamma y_{\rm cd} > y_{\rm cd}(J(\nu)=0)$, where constant
$\gamma \sim {\rm several}$, 
noting that $y_{\rm cd}({\rm H_2}; J(\nu)=0)$ is several times larger
than $y_{\rm cool}({\rm H_2})$ at $n \la n_{\rm cr}$.
This reduces to $\gamma k_{9} n_{\rm cr} > k_{31}$, or in another
expression, 
\begin{equation}
\label{eq:cd1}
J_{21} < 4 \times 10^{5} \alpha^{-1},
\end{equation}
where we have chosen $\gamma =5$.
The condition on $y_{\rm pd}({\rm H_2})$ leads to another constraint on
$J_{21}$: 
\begin{equation}
\label{eq:cd2}
J_{21} < 1 \times 10^{6} x_{4} \beta^{-1}.
\end{equation}
Here, we have used $y_{\rm cool} \simeq 10^{-6}$, since
the dependence of $y_{\rm pd}({\rm H_2})$ on $n$ is so strong that the
precise value of $y_{\rm cool}({\rm H_2})$ is not important.
Also, $k_{\rm form}=k_{32}$ has been used.
Combining the conditions (\ref{eq:cd1}) and (\ref{eq:cd2}), the
critical intensity $J_{\rm cr}$ above which the cloud never forms
sufficient H$_2$ is
\begin{equation}
\label{eq:jcr}
J_{{\rm cr},21}={\rm min}(4 \times 10^{5} \alpha^{-1}, 1 \times 10^{6}
x_{4} \beta^{-1}).
\end{equation}
Substituting the parameters $\alpha$ and $\beta$ into equation
(\ref{eq:jcr}), we see that the former constraint is more restrictive
than the latter for both two types of radiation discussed in the text. 

\section{Radiative Processes}
\subsection{Atomic Hydrogen Lines and Two-photon Emission}
We model an atomic hydrogen as a five level system.
The radiative cooling rate due to atomic hydrogen lines per unit volume
is given by  
\begin{equation}
\label{eq:Hcool}
\Lambda_{\rm H~lines}= \sum_{ul} h \nu_{ul} \beta_{{\rm esc},ul} A_{ul}
n_{u}({\rm H}) 
[S_{\rm H}(\nu_{ul})-J_{\rm cont}(\nu_{ul})]/S_{\rm H}(\nu_{ul}),
\end{equation}
where $n_{u}({\rm H})$ is the population density of atomic hydrogen in
the upper energy level $u$, 
$A_{ul}$ is the Einstein radiation probability for a
spontaneous downward transition, $\beta_{{\rm esc}, ul}$ is the
probability for a emitted line photon to escape without absorption, and
$h \nu_{ul}$ is the energy difference between the upper level $u$ and
the lower level $l$, and $J_{\rm cont}(\nu_{ul})$ is the mean intensity
of overlapping continuum at the line frequency $\nu_{ul}$.  
The source function $S_{\rm H}(\nu_{ul})$ is given by
\begin{equation}
S_{\rm H}(\nu_{ul})=\frac{2 h
\nu_{ul}^{3}}{c^{2}}(\frac{g_{u}n_{l}({\rm H})}{g_{l}n_{u}({\rm
H})}-1)^{-1}.
\end{equation} 
The cooling rate due to the two photon emission is  
\begin{equation}
\Lambda_{\rm 2ph}= h \nu_{21} \beta_{\rm esc,2ph} \Lambda_{21}
n_{2}({\rm H}),
\end{equation}
where $\Lambda_{21}$ is the spontaneous downward transition probability by 
the two-photon emission.
The escape probability $\beta_{\rm esc,2ph}={\rm exp}(-\tau_{\rm a})$, 
where $\tau_{\rm a}$ is the absorption optical depth owing to other
continuum processes (a1-a7 in Table 1; See \S B.3).
We take only account of absorption by other continuum processes, since 
our clouds do not become optically thick to the two-photon continuum. 

Relative population of each energy level is obtained from a solution of
the equations of the detailed balance between levels (e.g., de Jong,
Dalgarno, \& Boland 1980; Tielens \& Hollenbach 1985)
\begin{equation}
\label{eq:levelpop}
n_{i}({\rm H})\sum_{j \neq i}^{n} R_{ij}=\sum_{j \neq i}^{n} n_{i}({\rm
H}) R_{ji}, 
\end{equation}
where $n$ is the total number of transitions included.  
The transition rate $R_{ij}$ from level $i$ to level $j$ is given by   
\begin{equation}
\label{eq:trrate}
R_{ij} = \left\{
\begin{array}{ll}
 (A_{ij} \beta_{{\rm esc},ij} + \Lambda_{ij} \beta_{\rm esc,2ph})
(1+Q_{ij})+C_{ij} 
& \mbox{for $i>j$} \\
 (g_{j}/g_{i}) (A_{ji} \beta_{{\rm esc},ij}  + \Lambda_{ji} \beta_{\rm
esc,2ph}) Q_{ji}+C_{ij}
& \mbox{for $i<j$},
\end{array}
\right.
\end{equation}
where $C_{ij}$ is the collisional transition rate, and
 
\begin{equation}
Q_{ij}=\frac{c^2}{2 h \nu_{ij}^3} J_{\rm cont}(\nu_{ij}).
\end{equation}

The relative population within the first excited states (i.e.,$2p$ and
$2s$ states) is obtained from the statistical equilibrium 
(Spitzer \& Greenstein 1951),
\begin{equation}
\frac{n_{2s}}{n_{2p}}=\frac{g_{2s}}{g_{2p}}
(\frac{C_{2s2p}}{C_{2s2p}+A_{2s1s}}), 
\end{equation}
where the statistical weight $g_{2s}=2$, and $g_{2p}=6$, the radiative
transition rate by two photon emission $A_{2s1s}=8.23 (\rm s^{-1})$,
and the collisional transition rate between the levels
\begin{equation}
C_{2s2p}=6.21 \times 10^{-4} T^{-1/2} {\rm ln}(5.7T)
[1+\frac{0.78}{{\rm ln}(5.7T)}] n(\rm e)(\rm s^{-1}).
\end{equation}
Using $n_{2p}$ obtained above, we may write 
\begin{equation}
  A_{21}=\frac{n_{2p}}{n_{2}}A_{2p1s}
\end{equation}
where $A_{2p1s}=6.27 \times 10^{8} (\rm s^{-1})$.

We assume LTE within levels of the same
 principal quantum number for $n \ge 3$. 
Averaged over angular momentum quantum numbers,  
$A_{31}=5.575 \times 10^{7}, A_{41}=1.278 \times 10^{7},
A_{51}=4.125 \times 10^{6}, A_{32}=4.410 \times 10^{7},
A_{42}=8.419 \times 10^{6}, A_{52}=2.530 \times 10^{6},
A_{43}=8.986 \times 10^{6}, A_{53}=2.201 \times 10^{6},
A_{54}=2.699 \times 10^{6}$
(Janev et al. 1987).

Likewise, we use the two-photon emission rate
$\Lambda_{21}=(n_{2s}/n_{2})A_{2s1s}$ and 
$\Lambda_{ul}=0$ for other transitions. 

The collisional de-excitation rate is given by
\begin{equation}
C_{ul}=\gamma_{ul}(e) n(e) + \gamma_{ul}({\rm H}) n({\rm H}),
\end{equation}
where the collisional de-excitation rate coefficients $\gamma_{ul}(e)$
for collisions with electron and $\gamma_{ul}({\rm H})$ for those  
with atomic hydrogen are given below. 

The collisional de-excitation rate coefficients for collisions with
electron is given by 
\begin{equation}
  \gamma_{ul}(e)=10^{-8}(\frac{l^{2}}{u^{2}-l^{2}})^{3/2}
  \frac{l^{4}}{u^{2}} \alpha_{lu} \frac{\sqrt{\beta (\beta +
  1)}}{\beta + \chi_{lu}},~ 
\beta = \frac{h(\nu_{l}-\nu_{u})}{kT}, 
\end{equation}
(Sobelman et al. 1981), where $\alpha_{12}=24, \alpha_{13}=22,
\alpha_{14}=22, \alpha_{15}=21, \alpha_{23}=67, \alpha_{24}=58,
\alpha_{25}=56, \alpha_{34}=124, \alpha_{35}=101, \alpha_{45}=185$
and $\chi_{12}=0.28, \chi_{13}=0.37, \chi_{14}=0.39, \chi_{15}=0.41,
\chi_{23}=0.30, \chi_{24}=0.45, \chi_{25}=0.52, \chi_{34}=0.26,
\chi_{35}=0.42, \chi_{45}=0.21$.
For collisions with atomic hydrogen (Drawin 1969), 
\begin{equation}
  \gamma_{ul}({\rm H})=7.86 \times 10^{-15} (\frac{l}{u})^{2}
(1/l^{2}-1/u^{2})^{-2} 
f_{lu} T^{1/2} \frac{1+1.27 \times 10^{-5} (1/l^{2}-1/u^{2})^{-1} T}
{1+4.76 \times 10^{-17} (1/l^{2}-1/u^{2})^{-2} T^{2}},
\end{equation}
where $f_{12}=0.4162, f_{13}=7.910 \times 10^{-2}, f_{14}=2.899 \times
10^{-2}, f_{15}=1.394 \times 10^{-2}, f_{23}=0.6407, f_{24}=0.1193, 
f_{25}=4.467 \times 10^{-2}, f_{34}=0.8421, f_{35}=0.1506,
f_{45}=1.038$.

The collisional excitation rate can be obtained from the detailed balance:
\begin{equation}
C_{lu}=C_{ul}(g_{u}/g_{l}){\rm exp}(-h \nu_{ul}/kT).
\end{equation}

Taking into account the collisional de-excitation of line photons and
absorption due to the overlapping continuum, we may write the escape
probability as   
\begin{equation}
\beta_{{\rm esc},ul}=\frac{p_{ul}^{N_{\rm esc}}}{1+N_{\rm esc}} 
{\rm exp}(-\tau_{\rm a}),
\end{equation}
where $p_{ul}$ is the probability for an absorbed line photon to
re-emerge as an original line photon with neither collisional
de-excitation nor two photon continuum emission, and $N_{\rm esc}$ is the
number of scattering that an average photon experiences before escape.
The re-emergence probability can be written as
\begin{equation}
p_{ul}=\frac{A_{ul}}{A_{ul}+C_{ul}+\Lambda_{ul}}.
\end{equation}
The average number of scattering for Ly$\alpha$ line 
in large optical depth $\tau_{21}$ limit is given by
(Bonilha et al. 1979)  
\begin{equation}
\label{eq:Nesc_a}
N_{\rm esc}=1.6 \tau_{21} (1+0.05 \xi^{1.5})^{-1}
\end{equation}
for escape from a moving infinite slab, 
where $\xi=0.90 v_{\rm bulk}/v_{\rm D}$, $v_{\rm bulk}$ is the velocity
difference between the center and the edge of the slab, and
$v_{\rm D}=\sqrt{2kT/m_{\rm H}}$.
Here, as $v_{\rm bulk}$, we take the velocity at the edge of the
central core region whose radius is $R_{\rm c}$.
Then $v_{\rm bulk}=R_{\rm c}/3t_{ff}$.

For other atomic lines, we take  
\begin{equation}
N_{\rm esc} = \left\{
\begin{array}{ll}
 1.6 \tau_{ul}
& \mbox{for $\tau_{ul}<10^{4}$} \\
 160 \tau_{ul}^{1/2}
& \mbox{for $\tau_{ul}>10^{4}$}
\end{array}
\right.
\end{equation}
(e.g., Elitzur \& Ferland 1986).
The optical depth averaged over a line is given by
\begin{equation}
\tau_{ul}=\frac{A_{ul} c^{3}}{8 \pi
  \nu_{ul}^{3}}[n_{l}({\rm H})g_{u}/g_{l}-n_{u}({\rm H})]l_{\rm sh}/
  v_{\rm D}.  
\end{equation}
In order to take into account the effect of velocity gradient, we reduce
the shielding length $l_{\rm sh}$ for lines other than Ly$\alpha$ in
the case of large velocity gradient as   
\begin{equation}
l_{\rm sh}={\rm min}(R_{\rm c}, \Delta s_{\rm th})
\end{equation}
where
\begin{equation}
\Delta s_{\rm th}=v_{\rm D}/(\frac{dv}{dr})=3 v_{\rm D}
t_{\rm ff}.
\end{equation}
Recall that the correction for velocity gradient has been already included 
in the expression (\ref{eq:Nesc_a}) for Ly$\alpha$ photons.
Then we take $l_{\rm sh}=R_{\rm c}$ for the Ly$\alpha$ line.  

\subsection{Molecular Hydrogen Lines} 
The radiative cooling rate $\Lambda_{\rm H_2}$ due to molecular hydrogen
lines can be represented by the similar expression as
equation (\ref{eq:Hcool}).
We compute rovibrational population $n_{vJ}({\rm H_2})$ of H$_2$
following Hollenbach \& McKee (1979) using the renewed collisional rate
coefficients given by Hollenbach \& McKee (1989).
We take the spontaneous radiative decay rates $A_{vJ,v'J'}$ from Turner, 
Kirby-Docken, \& Dalgarno (1977).
Also we take into account the effects of the external radiation by the
same way as described for the atomic hydrogen lines (i.e., equation
\ref{eq:trrate}).   
The escape probability is given by (Takahashi, Hollenbach, \& Silk 1983)
\begin{equation}
\beta_{{\rm esc}, ul}=[(1-{\rm e}^{-\tau_{ul}})/ \tau_{ul}] {\rm e}^{-
\tau_{\rm a}}, 
\end{equation}
for the case that the velocity is proportional to the radius.
We consider the first three vibrational states ($v=0-2$) with
rotational levels up to $J=20$ in each vibrational state.
We assume the ortho to para ratio of molecular hydrogen to be the
equilibrium value 3:1.
We use the rovibrational energies $E(v,J)$ of Borysow, Frommhold, \&
Moraldi (1989).

\subsection{Compton Coupling}
The cooling rate of a gas with electron density $n_{\rm e}$ and
temperature $T$ embedded in a black body radiation field of energy
density $u$ and temperature $T_{\gamma}$ is given by (Kompaneets 1957;
Weymann 1965)
\begin{eqnarray}
\Lambda_{\rm Compt} &=& \frac{4 k (T-T_{\gamma})}{m_{e} c^{2}}c \sigma_{\rm
  T} n_{e} u ~~({\rm ergs~s^{-1} cm^{-3}}) \\
  &=& \frac{16k \sigma_{\rm T} \sigma_{\rm B} n_{e}}{m_{e} c^2}
  T_{\gamma}^4 (T-T_{\gamma}) ~~({\rm ergs~s^{-1} cm^{-3}}), 
\end{eqnarray}
where $\sigma_{\rm T}$ is the Thomson cross section, $\sigma_{\rm B}$ is 
the Stefan-Boltzmann constant, and $m_{e}$ is the mass of an
electron.
In order to treat the situation that the Compton energy transfer and
continuum processes coexist both in optically thin and thick cases, we
deal with the Compton process as if it was one of continuum sources, 
denoting 
\begin{equation}
\kappa_{\rm Compt}(\nu)=\frac{4kT_{\gamma}}{m_{e}c^{2}}\sigma_{\rm
  T}n_{e} = a_{\rm Compt} T_{\gamma},
\end{equation} 
and
\begin{equation}
\eta_{\rm Compt}(\nu)=\frac{4kT}{m_{e}c^{2}} \sigma_{\rm
  T}n_{e}J(\nu) = a_{\rm Compt} T J(\nu),
\end{equation}
where $a_{\rm Compt} \equiv (4k/m_{e}c^{2}) \sigma_{\rm T}n_{e}$.
This method reproduces the correct behavior both in optically thin and
  thick limit (Note that $T_{\gamma}=T$ in the optically thick limit).
In evaluating the radiation temperature in the above equation, we use
the definition
\begin{equation}
T_{\gamma} \equiv (\pi J/ \sigma_{\rm B})^{1/4},
\end{equation}
where the mean intensity $J=\int J(\nu) d\nu$.

\subsection{Continuum}
We consider the following continuum processes (cf., Lenzuni et al. 1991).
Among pure absorption, bound-free absorption of ${\rm H, He, H^{-}, H_2^{+}}$,
free-free absorption of ${\rm H^{-}, H}$, H$_2$ collision-induced
absorption,
\footnote{Two-photon emission of atomic hydrogen was treated as
  described in \S A.1.}
and among scattering, Rayleigh scattering of H, and Thomson scattering are
  included. 
The cross sections are listed in Table 1.

The net rate of energy transport from matter to radiation per unit
volume per unit frequency is 
\begin{equation}
\label{eq:cnt1}
\Lambda_{\rm cont}(\nu)=4 \pi [\eta_{\rm a}(\nu)-\kappa_{\rm a}(\nu)
J(\nu)],
\end{equation}
where $\eta_{\rm a}(\nu)$ is the thermal part of the emission coefficient
and $\kappa_{\rm a}(\nu)$ is the true absorption coefficient.  
Also, we use the scattering coefficient $\kappa_{\rm s}(\nu)$ below. 
Specifically,
\begin{equation}
\eta_{\rm a}(\nu)=\sum_{\rm aj=a1}^{\rm a8} \eta_{\rm aj}(\nu)
\end{equation}
and
\begin{equation}
\kappa_{\rm a}(\nu)=\sum_{\rm aj=a1}^{\rm a8} \kappa_{\rm aj}(\nu),~~
\kappa_{\rm s}(\nu)=\kappa_{\rm s1}(\nu)+\kappa_{\rm s2}(\nu).
\end{equation}

We assume that the radiation field is static inside the cloud, then
\begin{equation}
\label{eq:cnt2}
\Lambda_{\rm cont}(\nu)=\frac{u(\nu)-u_{\rm ex}(\nu)}{t_{\rm esc}(\nu)},
\end{equation}
where $u(\nu)$ and $u_{\rm ex}(\nu)$ are the radiation energy density in 
side and outside the cloud.
The photon escape time
\begin{equation}
t_{\rm esc}(\nu)=\frac{x(\nu)}{\chi(\nu) c},  
\end{equation}
where the extinction coefficient
\begin{equation}
\chi(\nu)=\kappa_{\rm a}(\nu)+\kappa_{\rm s}(\nu),
\end{equation}
and
\begin{equation}
x(\nu) \equiv {\rm max}[\tau(\nu)^2,\tau(\nu)],
\end{equation}
by using optical depth $\tau(\nu)=\chi(\nu)R_{\rm c}$.
Equating equations (\ref{eq:cnt1}) and (\ref{eq:cnt2}), and using 
$u(\nu)=4 \pi J(\nu)/c$, we obtain 
\begin{equation}
\label{eq:J1}
J(\nu)=\frac{J_{\rm ex}(\nu) + \xi (\nu) x (\nu) S_{\rm a}(\nu)}{1 +
\xi (\nu) x (\nu)}, 
\end{equation}
where $\xi (\nu) \equiv \kappa_{\rm a}(\nu)/ \chi (\nu)$.
The external radiation field $J_{\rm ex}(\nu)=J_{\rm
UV}(\nu)+B(\nu;T_{\rm CMBR})$ in our case, where $T_{\rm CMBR}$ is the
temperature of the CMBR. 
If we include the Compton energy transfer, substituting 
\begin{equation}
\frac{\eta_{\rm a}(\nu)+\eta_{\rm Compt}(\nu)}{\kappa_{\rm a}(\nu)+\kappa_{\rm
Compt}(\nu)}=\frac{\eta_{\rm a}(\nu)+a_{\rm Compt}T J(\nu)}{\kappa_{\rm
a}(\nu)+a_{\rm Compt} T_{\gamma}}
\end{equation}
into equation (\ref{eq:J1}) instead of $S_{\rm a}(\nu)$, we obtain the
final expression for the mean intensity inside the cloud: 
\begin{equation}
\label{eq:cnt3}
J(\nu)=\frac{J_{\rm ex}(\nu) + \xi (\nu) x(\nu) \eta_{\rm
  a}(\nu)/(\kappa_{\rm a}(\nu)+a_{\rm Compt} T_{\gamma})}
{1 + \xi (\nu) x (\nu)[1 - a_{\rm Compt}T/(\kappa_{\rm a}(\nu)+a_{\rm
  Compt} T_{\gamma})]}.
\end{equation}

Substituting equation (\ref{eq:cnt3}) into (\ref{eq:cnt1}) and integrating
over frequency $\nu$, we obtain the sum of cooling rates $\Lambda_{\rm
cont}+\Lambda_{\rm Compt}$ due to the continuum and the Compton
coupling. 

In the following, we describe how to calculate the emission and absorption
coefficients using the cross sections. 

\subsubsection{bound-free absorption / free-bound emission {\rm (a1-a4)}}
Consider the radiative association RA($i$) of species A and B into $i$th state
of C, whose binding energy is $h \nu_{i}$, and resultant free-bound emission,
\begin{equation}
{\rm RA({\it i}):  A + B \rightarrow C({\it i}) + {\it h \nu}},
\end{equation}
and its inverse reaction PD({\it i})(i.e., photodissociation of C from
$i$th state, or the bound-free absorption).
The cross section of this (and its inverse) reaction is $\sigma_{{\rm
  RA} i}(v) (\sigma_{{\rm PD} i}(\nu)$, respectively). 
In our case, the species C means H, He, H$^{-}$, and H$_2^{+}$ for the
processes a1-a4, respectively.

From the Milne relation (e.g., Rybicki \& Lightman 1979), 
\begin{equation}
\sigma_{{\rm RA} i}=\sigma_{{\rm PD} i} (\frac{h \nu}{m_{\rm r} c v})^{2}
\frac{2g_{{\rm C} i}}{z_{\rm A}z_{\rm B}},
\end{equation}
where the reduced mass $m_{\rm r}=m_{\rm A}m_{\rm B}/m_{\rm C}$,
$z_{\rm A}(z_{\rm B})$ is the partition function of A (B,
respectively), $g_{{\rm C} i}$ is the statistical weight of the $i$th
state of C, and 
\begin{equation}
\frac{1}{2}m_{\rm r}v^{2}=h(\nu-\nu_i).
\end{equation}

The emission coefficient of RA($i$) is given by 
\begin{eqnarray}
\eta_{{\rm RA} i}(\nu)&=&\frac{h \nu}{4 \pi} \sigma_{{\rm RA} i} 
n({\rm A}) n({\rm B}) vf(v) \frac{dv}{d \nu}\\
&=&\frac{2h \nu^{3}}{c^2} \frac{g_{{\rm C} i}}{z_{\rm A}z_{\rm B}}
(\frac{h^2}{2 \pi m_{\rm r}kT})^{3/2}  
\sigma_{{\rm PD} i}
{\rm exp}(- \frac{h (\nu -\nu_i)}{kT}) n({\rm A}) n({\rm B}), 
\end{eqnarray}
where the distribution function of the relative velocity between particles 
A and B 
\begin{equation}
f(v)=4 \pi (\frac{m_{\rm r}}{2 \pi kT})^{3/2} 
{\rm exp}(-\frac{m_{\rm r}v^2}{2kT}) v^{2}.
\end{equation}
Also, the absorption coefficient of PD($i$) is given by 
\begin{eqnarray}
\kappa_{{\rm PD} i}(\nu)&=&\sigma_{{\rm PD} i} n_{i}({\rm C})
-\frac{h \nu}{4 \pi} \frac{c^2}{2h \nu^{3}}
\sigma_{{\rm RA} i}n({\rm A})n({\rm B})vf(v)\frac{dv}{d \nu}\\
&=&\sigma_{{\rm PD} i} n_{i}({\rm C})
-\frac{g_{{\rm C} i}}{z_{\rm A}z_{\rm B}} 
(\frac{h^2}{2 \pi m_{\rm r} kT})^{3/2} \sigma_{{\rm PD} i} 
{\rm exp}(- \frac{h(\nu-\nu_i)}{kT}) n({\rm A})n({\rm B}).
\end{eqnarray}
The second term in the equation above represents the induced
association.

Summing over all levels, we obtain the emission and absorption
coefficient owing to the reaction 
\begin{equation}
{\rm RA:  A + B \rightarrow C + {\it h \nu}},
\end{equation}
and its inverse PD;
\begin{eqnarray}
\eta_{\rm RA} (\nu)&=&\sum_{i} \eta_{i}(\nu)\\
&=& \frac{2h \nu^{3}}{c^2} \sigma_{\rm PD}^{\ast} n({\rm C}) 
{\rm exp}(-\frac{h \nu}{kT}) [ K(T)^{-1} \frac{n({\rm A}) n({\rm
    B})}{n({\rm C})} ],
\end{eqnarray}
and
\begin{eqnarray}
\kappa_{\rm PD} (\nu) &=& \sum_{i} \kappa_{i}(\nu)\\
&=& \sigma_{\rm PD}n({\rm C})\{1-{\rm exp}(-\frac{h \nu}{kT})
[\frac{\sigma_{\rm PD}^{\ast}}{\sigma_{\rm PD}}K(T)^{-1} \frac{n({\rm A})
  n({\rm B})}{n({\rm C})} ] \}.
\end{eqnarray}
In the above equations, the equilibrium constant 
\begin{eqnarray}
K(T) & \equiv &[\frac{n({\rm A})n({\rm B})}{n({\rm C})}]^{\ast}\\
&=&\frac{z_{\rm A}z_{\rm B}}{z_{\rm C}}(\frac{2 \pi m_{\rm r}kT}{h^2})^{3/2}
{\rm exp}(- \frac{h \nu_1}{kT}),
\end{eqnarray}
and
\begin{equation}
\sigma_{\rm PD}=\sum_{i} c_{i}\sigma_{{\rm PD} i},~~
\sigma_{\rm PD}^{\ast}=\sum_{i} c_{i}^{\ast} \sigma_{{\rm PD} i},
\end{equation}
where the relative population of level $i$  is
 $c_{i}=n_{i}({\rm C})/n({\rm C})$,
and its LTE value 
$c_{i}^{\ast}=g_{{\rm C} i} {\rm exp}(- h
    (\nu_1-\nu_i)/kT)/z_{\rm C}$.

\subsubsection{free-free absorption/ emission {\rm (a5,a6)} and H$_2$
collision-induced absorption/ emission {\rm (a7,a8)} } 
First, we consider the free-free emission and absorption (a5,a6)
\begin{equation}
{\rm FF:  A + {\it e} \leftrightarrow A + {\it e} + \gamma},
\end{equation}
where A=H, and H$^{+}$ for the processes a5, and a6, respectively.
The absorption cross section for A is $\sigma_{\rm FF}$. 
Note that $\sigma_{\rm FF} \propto n(e)$ (see Table 1).
This process is collisional, then it occurs at the LTE rate.
Thus, 
\begin{equation}
\eta_{\rm FF} (\nu)=\frac{2h \nu^{3}}{c^2} \sigma_{\rm FF} n({\rm A}) 
{\rm exp}(-\frac{h \nu}{kT}),
\end{equation}
and
\begin{equation}
\kappa_{\rm FF} (\nu)=\sigma_{\rm FF}n({\rm A})\{1-{\rm exp}(-\frac{h
\nu}{kT})\}. 
\end{equation}

Next, we consider the H$_2$ collision-induced emission and absorption
(a7,a8)
\begin{equation}
{\rm CI:  H_2 + B \leftrightarrow H_2 + B + \gamma},
\end{equation}
In this equation, B means H$_2$ for the process a7, and He for a8.
The absorption cross section for H$_2$ is $\sigma_{\rm CI}$. 
This process occurs only at so high density that 
$\sigma_{\rm CI}=\sigma_{\rm CI}^{\ast}$.
Then, in the same way as the above,    
\begin{equation}
\eta_{\rm CI} (\nu)=\frac{2h \nu^{3}}{c^2} \sigma_{\rm CI} n({\rm H_2}) 
{\rm exp}(-\frac{h \nu}{kT}),
\end{equation}
and
\begin{equation}
\kappa_{\rm CI} (\nu)=\sigma_{\rm CI}n({\rm H_2})\{1-{\rm exp}(-\frac{h
\nu}{kT})\}. 
\end{equation}

\subsubsection{scattering {\rm (s1,s2)}}
We denote the cross section of A for the scattering (s1,s2)
\begin{equation}
{\rm SC:  A + \gamma \leftrightarrow A + \gamma '}
\end{equation}
as $\sigma_{\rm SC}(\nu)$.
In the above expression, A=H, and {\it e} for the scattering s1, and s2,
respectively.
The scattering coefficient is given by
\begin{equation}
\kappa_{\rm SC} (\nu)=\sigma_{\rm SC}(\nu) n({\rm A}).
\end{equation} 

\section{Chemical Reactions}
We treat non-equilibrium chemistry of hydrogen-helium gas between the
following nine species: ${\rm
H,H_{2},{\it e},H^{+},H_{2}^{+},H^{-},He,He^{+}}$, and ${\rm He^{++}}$. 
Included reactions and their rate coefficients are presented in Table 2.

When the gas density rises, and consequently the chemical reaction
timescale becomes shorter than the collapse timescale 
of the cloud, the chemical equilibrium is reached.
In addition to this, if the cloud becomes optically thick to the
continuum, and as a result, the radiation field reduces to the black
body radiation of the matter temperature, the chemical equilibrium
reduce to the Saha value.     
In order to reproduce this feature, inverse processes are included for
all reactions.    
Rates of inverse reactions are computed from the principle of detailed
balance.

We switch to equilibrium chemistry at a sufficiently high density
$n_{\rm eq}$ where Saha equilibrium has been already reached.
In our case, $n_{\rm eq}=10^{18-19} {\rm cm^{-3}}$.

The rate coefficients of the radiative association
\begin{equation}
{\rm RA:  A + B \rightarrow C + \gamma},
\end{equation}
and its inverse reaction PD, i.e. the photodissociation of C, can be
written as   
\begin{equation}
\label{eq:kra}
k_{\rm RA}=K(T)^{-1} \int_{0}^{\infty} \frac{4 \pi B(\nu; T)}{h \nu}
\sigma_{\rm PD}^{\ast}[1-{\rm exp}(- \frac{h \nu}{kT})] d \nu,
\end{equation}
and 
\begin{equation}
\label{eq:kpd}
k_{\rm PD}=\int_{0}^{\infty} \frac{4 \pi J(\nu)}{h \nu} 
\sigma_{\rm PD} \{1-{\rm exp}(- \frac{h \nu}{kT})[\frac{\sigma_{\rm
    PD}^{\ast}}{\sigma_{\rm PD}}K(T)^{-1} \frac{n({\rm A})n({\rm
B})}{n({\rm C})}] \} d \nu.
\end{equation}
from the discussion in \S A 4.1. 

The partition function of H
\begin{equation}
z_{\rm H}=\sum_{n=1}^{5} g_{n} {\rm exp}(-E_{n}/kT)
\end{equation} 
where $g_{n}=2n^{2}$ and $E_{n}=13.598{\rm eV}/n^{2}$.
For H$_2$, the sum extends over $0 \le v \le 5$ and $0 \le J \le 25$. 
The rovibrational energies $E(v,J)$ are taken from Borysow et al.(1989).
The partition function of ${\rm H_{2}^{+}}$ is taken from Patch \&
McBride (1968).
For other species, we take $z_{e}=2$, $z_{\rm H^{+}}=1$, 
$z_{\rm H^{-}}=1$, $z_{\rm He}=1$, $z_{\rm He^{+}}=2$, and $z_{\rm
He^{++}}=1$.
 
%%%%%%%%%%%%%%%%%%%%%%%%%%%%%%%%%%%%%%%%%%%%%%%%%%%%%%%%%%%%%%%%%%
%%%%%% (6) References   %%%%%%%%%%%%%%%%%%%%%%%%%%%%%%%%%%%%%%%%%%
%%%%%%%%%%%%%%%%%%%%%%%%%%%%%%%%%%%%%%%%%%%%%%%%%%%%%%%%%%%%%%%%%%

\clearpage

\begin{deluxetable}{clllll}
\tablecaption{Continuum Processes
\label{tab:cont} }
\tablehead{
\colhead{number} &
\colhead{name} &
\colhead{process} & 
\colhead{cross section (cm$^{2}$)} & 
\colhead{reference} & 
} 
\startdata
a1& H b-f
&  ${\rm H({\it n})   +  \gamma   \rightarrow   H^{+}   +  {\it e}}$
&  $7.909 \times 10^{-18} n (\nu/\nu_{n})^{-3}$
&  1 \\
&
&
& ~$h\nu_{n}=13.598{\rm eV}/n^{2}$  
&  \\
a2& He b-f
&  ${\rm He  +  \gamma   \rightarrow   He^{+}  +  {\it e}}$
&  $7.83 \times 10^{-18}[1.66(\nu/\nu_{\rm T})^{-2.05}-0.66(\nu/\nu_{\rm
  T})^{-3.05}]$,
&  2 \\
&
&
& ~$h\nu_{\rm T}=24.586{\rm eV}$  
&  \\
a3& H$^{-}$ b-f
& ${\rm H^{-}  +  \gamma   \rightarrow   H   +  {\it e}}$
&  $10^{-18} \lambda^{3} (1/ \lambda - 1/ \lambda_{0})^{3/2} f(\lambda),~
\lambda_0=1.6419 {\rm \mu m}$
&  3 \\
&
&
& ~$f(\lambda)$ given by Eq.(5) of ref.
& \\
a4& H$_{2}^{+}$ b-f
& ${\rm H_{2}^{+}  +  \gamma   \rightarrow   H   +  H^{+}}$
&  see Table 2 of ref.
&  4 \\
a5& H$^{-}$ f-f 
&  ${\rm H  +  {\it e} + \gamma   \rightarrow   H   +  {\it e}}$
&  $k_{\lambda}^{\rm ff}(T) k_{\rm B}T n_{e}$
&  3 \\
&
&
&  ~$k_{\lambda}^{\rm ff}(T)$ given by Eq.(6) of ref. 
& \\
a6& H f-f 
&  ${\rm H^{+} + {\it e} + \gamma  \rightarrow   H^{+}   + {\it e}}$
&  $3.692 \times 10^{8} g_{\rm ff}(\nu,T) \nu^{-3} T^{-1/2} n_{e}$,
&  \\
& 
&
&  ~we take $g_{\rm ff}(\nu,T)=1$ 
&  1  \\
a7& H$_2$-H$_2$ CIA
&  ${\rm H_2({\it v,J}) + H_2 + \gamma \rightarrow H_2({\it v',J'}) + H_2 }$
&   see Fig.1 of ref.
&  5  \\
a8& H$_2$-He CIA
&  ${\rm H_2({\it v,J}) + He + \gamma \rightarrow H_2({\it v',J'}) + He }$
&   see Fig.2 of ref.
&  5  \\
s1& H Rayleigh
&  $ {\rm H + \gamma \rightarrow H + \gamma'}$
&  $5.799 \times 10^{-29} \lambda^{-4} + 1.422 \times 10^{-30} \lambda^{-6}$
&  6 \\
&
&
&  ~$+2.784 \times 10^{-32} \lambda^{-8}$
&  \\
s2& Thomson  
&  $ {\rm {\it e} + \gamma \rightarrow {\it e} + \gamma'}$
&  $6.65 \times 10^{-25}$
&  1 \\
\enddata
\tablecomments{The wavelength $\lambda$ is in $\mu$m.}
\tablerefs{(1) Rybicki \& Lightman (1979), 
(2) Osterbrock (1989), (3) John (1988), (4) Stancil (1994),  
(5) Borysow et al. (1997), (6) Kurucz (1970)}
\end{deluxetable}

\begin{deluxetable}{clllr}
\tablecaption{Chemical Reactions
\label{tab:react} }
\tablehead{
\colhead{Number} &
\colhead{Reaction} & 
\colhead{Rate Coefficient} & 
\colhead{Reference} & 
} 
\startdata
1,2& ${\rm H({\it n}) + {\it e} \rightleftharpoons H^{+} + 2 {\it e}}$
& $k_{1}=\sum_{n=1}^{5} c_{n} k_{1,n}$
& \\
&
& ~~$k_{1,n=1}=4.25 \times 10^{-11} T^{1/2} {\rm exp}(- \chi_H/ kT)$
& 1 \\
& 
& ~~$k_{1,n=2}=6.69 \times 10^{-10} T^{1/2} {\rm exp}(- \chi_H/2^{2}/kT)$ 
  ($T<$4680K)
& 1 \\
& 
& ~~$k_{1,n=2}$: see Janev et al. [2.1.6]  
  ($T>$4680K)
&  \\
& 
& ~~$k_{1,n \geq 3}=9.56 \times 10^{-6} T({\rm eV})^{-1.5}$
& 2 \\
&
& ~~~~~~~~~~~~$( \beta_{n}^{2.33}+4.38 \beta_{n}^{1.72}+1.32
\beta_{n})^{-1} {\rm exp}(-\beta_{n})$, 
& \\
&
& ~~~~~~~~$\beta_{n}=(\chi_{\rm H}/n^{2})/kT$
& \\
&
& $k_{2}=k_{1}^{\ast} (z_{\rm H}/z_{\rm H^{+}} z_{e})
   4.1414 \times 10^{-16} T^{-1.5} {\rm exp}(\chi_{\rm H}/kT)$
& \\
&
& ~~$k_{1}^{\ast}=\sum_{n=1}^{5} c_{n}^{\ast} k_{1,n}$
& \\
&
&
&
& \\
3,4& ${\rm H({\it n}) + H \rightleftharpoons H^{+} + {\it e} + H}$  
& $k_{3}=\sum_{n=1}^{5} c_{n} k_{3,n}$
& \\
&
& ~~$k_{3,1}=1.2 \times 10^{-17} T^{1.2} ~{\rm exp}(- \chi_{\rm H} /kT)$
& 1\\
&
& ~~$k_{3,n \ge 2}=7.86 \times 10^{-15} n^{4} f_{n} T^{0.5} 
(1 +1.27 \times 10^{-5} n^{2} T)$
& 3\\
&        
& ~~~~~~~~~~$(1 +4.76 \times 10^{-17} n^{4} T^{2})^{-1}
{\rm exp}(-\chi_{\rm H}/n^{2}/kT)$
& \\
&
&  ~~~~~~$f_{2}=0.71,~~f_{3}=0.81,~~f_{4}=0.94$
& \\
&
& $k_{4}=k_{3}^{\ast} (z_{\rm H}/z_{\rm H^{+}} z_{e})
   4.1414 \times 10^{-16} T^{-1.5} {\rm exp}(\chi_{\rm H}/kT)$
& \\
&
& ~~$k_{3}^{\ast}=\sum_{n=1}^{5} c_{n}^{\ast} k_{3,n}$
& \\
&
&
&
& \\
5,6&   ${\rm H^{-}  +   H^{+}   \rightleftharpoons H({\it n})+H }$ 
& $k_{5}=\sum_{n=1}^{3} k_{5,n}$
& \\
& 
& ~~$k_{5,1}=6.92 \times 10^{-14} T^{0.5}$
& 1\\
&
& ~~$k_{5,2}=8.0 \times 10^{-13} T^{0.83}$
& \\
&
& ~~$k_{5,3}=6.18 \times 10^{-7} T^{-0.27}$
& \\
& 
& $k_{6}=[\sum_{n=1}^{3} (c_{n}/c_{n}^{\ast}) k_{5,n}]$
& \\
&
& ~~~~~~~~~$(z_{\rm H^{-}}z_{\rm H^{+}}/{z_{\rm H}}^{2})
   {\rm exp}[-(\chi_{\rm H}- \chi_{\rm H^{-}})/kT]$
& \\
&
&
&
& \\
7,8&   ${\rm H_2  +   H^{+}   \rightleftharpoons   H    +   H_2^{+}  }$ 
& $k_{7}=1.5 \times 10^{-10} {\rm exp}(-1.4 \times
  10^{4}/T)~~~(T>10^{4}{\rm K})$ 
& 4 \\
& 
& $k_{7}=3 \times 10^{-10} {\rm exp}[- (\chi_{\rm H_2}-\chi_{\rm
{H_2}^{+}})/kT]~~~(T<10^{4}{\rm K})$ 
& \\
&
& $k_{8}=k_{7} (z_{\rm H^{+}}z_{\rm H_{2}}/z_{\rm H} z_{\rm {H_{2}}^{+}})
 {\rm exp}[(\chi_{\rm H_2}-\chi_{\rm {H_2}^{+}})/kT]$
& \\
&
&
&
& \\
9,10&   ${\rm H   +   H^{-}   \rightleftharpoons   H_2 +   {\it e}  }$  
&  $k_{9}=1.5 \times 10^{-9}~~~(T < 3 \times 10^{2} {\rm K})$
& 4 \\
&
&  $k_{9}=4.0 \times 10^{-9} T^{-0.17}~~~(T > 3 \times 10^{2} {\rm K})$
& \\
&  
&  $k_{10} = k_{10,\rm H}^{1-a} k_{10,\rm L}^a$
&    \\
&
&  ~~$k_{10,\rm L}=2.7 \times 10^{-8} T^{-1.27} 
{\rm exp}[-(\chi_{\rm H_{2}}-\chi_{\rm H^{-}})/kT]$
& 4\\
& 
&  ~~$k_{10,\rm H}=k_{9}(z_{\rm H} z_{\rm H^{-}}/z_{\rm H_2} z_{e})
    2.775 \times 10^{4} {\rm exp}[-(\chi_{\rm H_{2}}-\chi_{\rm
        H^{-}})/kT]$
&  \\
&
& ~~$a = (1+n/n_{\rm cr})^{-1}$
&    \\
&
& ~~${\rm log}_{10}(n_{\rm cr}) = 4.0-0.416 {\rm log}_{10}(T/10^4)$
&  5\\
&
& ~~~~~~~~~~~~~~~~~~$-0.327 ({\rm log}_{10}(T/10^4))^2$
&   \\
&
&
&
& \\
11,12&   ${\rm H_2^{+} + {\it e} \rightleftharpoons H({\it n}) + H  }$ 
&  $k_{11}=\sum_{n=1}^{5} k_{11,n}=2 \times 10^{-7} T^{-1/2}$
&  4\\
&
&  ~~$k_{11,1}:k_{11,2}:k_{11,3}:k_{11,4}:k_{11,5}=0:0.10:0.45:0.22:0.12$
&  2\\
&  
&  ~~~but, $k_{11,n}=0$  in case of $\chi_{\rm H}-\chi_{\rm
H_{2}^{+}}-\chi_{\rm H}/n^{2}<k T$  
&  \\
&
&  $k_{12}=[\sum_{n=1}^{5} (c_{n}/c_{n}^{\ast}) k_{11,n}]$ 
&  \\
&
& ~~~~~~~~~$(z_{\rm {H_2}^{+}}z_{e}/{z_{\rm H}}^{2}) 
3.6034 \times 10^{-5}{\rm exp}[-(\chi_{\rm H}-\chi_{\rm {H_2}^{+}})/kT]$
&  \\
&
&
&
& \\
13,14& ${\rm 2 H_2   \rightleftharpoons 2 H    +   H_2}$ 
&$k_{13} = k_{13,{\rm H}}^{1-a} k_{13,{\rm L}}^a$
& 5 \\
&
& ~~~~$k_{13,{\rm L}} = 1.18 \times 10^{-10} {\rm exp}(-6.95 \times 10^{4}/T)$
&  \\
&
& ~~~~$k_{13,{\rm H}} = 1.30 \times 10^{-9} {\rm exp}(-5.33 \times 10^{4}/T)$
&   \\
&
& ~~~~$a = (1+n/n_{\rm cr})^{-1}$
&    \\
&
& ~~~~${\rm log}_{10}(n_{\rm
  cr}) = 4.845-1.3 {\rm log}_{10}(T/10^4)$
&  \\
&
& ~~~~~~~~~~~~~~~~~~$+1.62 ({\rm log}_{10}(T/10^4))^2$
&   \\
&
&   $k_{14}=k_{13,{\rm H}} (z_{\rm H_2}/{z_{\rm H}}^{2}) 1.493 \times 10^{-20}
     T^{-1.5} {\rm exp}(\chi_{\rm H_2}/kT)$
&   \\
&
&
&
& \\
15,16&   ${\rm H_2  +   H    \rightleftharpoons 3 H  }$    
&  $k_{15} = k_{15,\rm H}^{1-a} k_{15,\rm L}^a$
&   \\
&
& ~~~~$k_{15,\rm L}=1.12 \times 10^{-10} {\rm exp}(-7.035 \times 10^{4}/T)$
&  5 \\
&
& ~~~~$k_{15,\rm H}=6.5 \times 10^{-7} T^{-1/2}$
&  6 \\
&
& ~~~~~~~~~~~${\rm exp}(- \chi_{\rm H_2}/kT)[1-{\rm exp}(-6000/T)]$
&    \\
&
& ~~~~$a = (1+n/n_{\rm cr})^{-1}$
&    \\
&
& ~~~~${\rm log}_{10}(n_{\rm
  cr}) = 4.0-0.416 {\rm log}_{10}(T/10^4)$
&  5 \\
&
& ~~~~~~~~~~~~~~~~~~$-0.327 ({\rm log}_{10}(T/10^4))^2$
&   \\
&
&  $k_{16}=k_{15,{\rm H}} (z_{\rm H_2}/{z_{\rm H}}^{2}) 1.493 \times 10^{-20}
     T^{-1.5} {\rm exp}(\chi_{\rm H_2}/kT)$
&   \\
&
&
&
& \\
17,18&   ${\rm H_2  + {\it e} \rightleftharpoons 2 H    +   {\it e}  }$ 
&  $k_{17}=1.3 \times 10^{-18} T^{2} {\rm exp}(- \chi_{\rm H_2}/kT)$
& 1 \\
&
&  $k_{18}=k_{17} (z_{\rm H_2}/{z_{\rm H}}^{2}) 1.493 \times 10^{-20}
     T^{-1.5} {\rm exp}(\chi_{\rm H_2}/kT)$
& \\
&
&
&
& \\
19,20&   ${\rm H^{-}  + {\it e} \rightleftharpoons   H    + 2 {\it e}}$ 
&  $k_{19}=4 \times 10^{-12} T {\rm exp}(-\chi_{\rm H^{-}}/kT)$
& 5 \\
&
&  $k_{20}=k_{19} (z_{\rm H^{-}}/z_{\rm H} z_{e}) 
 4.1414 \times 10^{-16} T^{-1.5} {\rm exp}(\chi_{\rm H^{-}}/kT)$
& \\
&
&
&
& \\
21,22&   ${\rm H_2^{+} + {\it e} \rightleftharpoons   H + H^{+}  +
  {\it e}  }$ 
&  $k_{21}=2 \times 10^{-7} {\rm exp}(- \chi_{\rm H_2^{+}}/kT)$
& 1 \\
&  
&  $k_{22}=k_{21}(z_{\rm H_2^{+}}/z_{\rm H} z_{\rm H^{+}}) 
     1.493 \times 10^{-20} T^{-1.5} {\rm exp}(\chi_{\rm H_2^{+}}/kT)$
& \\
&
&
&
& \\
23,24&   ${\rm He  + {\it e} \rightleftharpoons   He^{+}  + 2 {\it e} }$ 
&  $k_{23}=2.38 \times 10^{-11} T^{1/2} [1+(T/10^{5})]^{-1} 
{\rm exp}(-\chi_{\rm He}/kT)$
& 7 \\
&  
&  $k_{24}=k_{23}(z_{\rm He}/z_{\rm {He}^{+}}z_{e})
4.1414 \times 10^{-16} T^{-1.5} {\rm exp}(\chi_{\rm He}/kT)$
& \\
&
&
&
& \\
25,26&  ${\rm He^{+} + {\it e}  \rightleftharpoons He^{++} +  2 {\it e} }$ 
&  $k_{25}=5.68 \times 10^{-12} T^{1/2} [1+(T/10^{5})]^{-1} 
{\rm exp}(-\chi_{\rm He^{+}}/kT)$
& 7 \\
&
&  $k_{26}=k_{25} (z_{\rm {He}^{+}}/z_{\rm {He}^{++}} z_{e})
4.1414 \times 10^{-16} T^{-1.5} {\rm exp}(\chi_{\rm {He}^{+}}/kT)$
& \\
&
&
&
& \\
27,28&   ${\rm H   +   \gamma  \rightleftharpoons   H^{+}   +  {\it e} }$ 
& see equations (\ref{eq:kra}), (\ref{eq:kpd}) in the text
& \\
& 
& and process (a1) in Table 1
& \\
&
&
&
& \\
29,30&   ${\rm He  +   \gamma  \rightleftharpoons   He^{+}  +  {\it e} }$ 
& see equations (\ref{eq:kra}), (\ref{eq:kpd}) in the text
& \\
&
& and process (a2) in Table 1
& \\
&
&
&
& \\
31,32&   ${\rm H^{-}  +   \gamma  \rightleftharpoons   H    +  {\it e} }$ 
& see equations (\ref{eq:kra}), (\ref{eq:kpd}) in the text
& \\
&
& and process (a3) in Table 1
& \\
&
&
&
& \\
33,34&   ${\rm H_2^{+} +   \gamma  \rightleftharpoons   H    +   H^{+} }$ 
& see equations (\ref{eq:kra}), (\ref{eq:kpd}) in the text
& \\
& 
& and process (a4) in Table 1
& \\
&
&
&
& \\
35&   ${\rm H_2 +  \gamma  \rightarrow H_2^{\ast} \rightarrow 2 H}$ 
& $k_{35}=1.4 \times 10^{9} J(h \nu=12.4 {\rm eV}) f_{\rm sh}$
& 8 \\
&
& ~~~~~$f_{\rm sh}={\rm min} \left[ 1,(N_{\rm H_2}/10^{14}{\rm
cm^{-2}})^{-3/4} \right]$ 
& \\
\enddata
\tablecomments{The temperature $T$ is in K, except otherwise noted.}
\tablecomments{The binding energies are
$\chi_{\rm H}/k=1.578 \times 10^{5}$K,
$\chi_{\rm H_2}/k=5.197 \times 10^{4}$K,
$\chi_{\rm H^{-}}/k=8.761 \times 10^{3}$K,
$\chi_{\rm H_2^{+}}/k=3.067 \times 10^{4}$K,
$\chi_{\rm He}/k=2.853 \times 10^{5}$K,
and
$\chi_{\rm He^{+}}/k=6.312 \times 10^{5}$K.
}
\tablerefs{(1) Lenzuni et al. (1991), (2) Janev et al. (1987), (3)
Drawin (1969), (4) Galli \& Palla (1998), (5) Shapiro \& Kang (1987),
(6) Palla et al. (1983), (7) Black (1981), (8) Draine \& Bertoldi
(1996)}
\end{deluxetable}

\figcaption[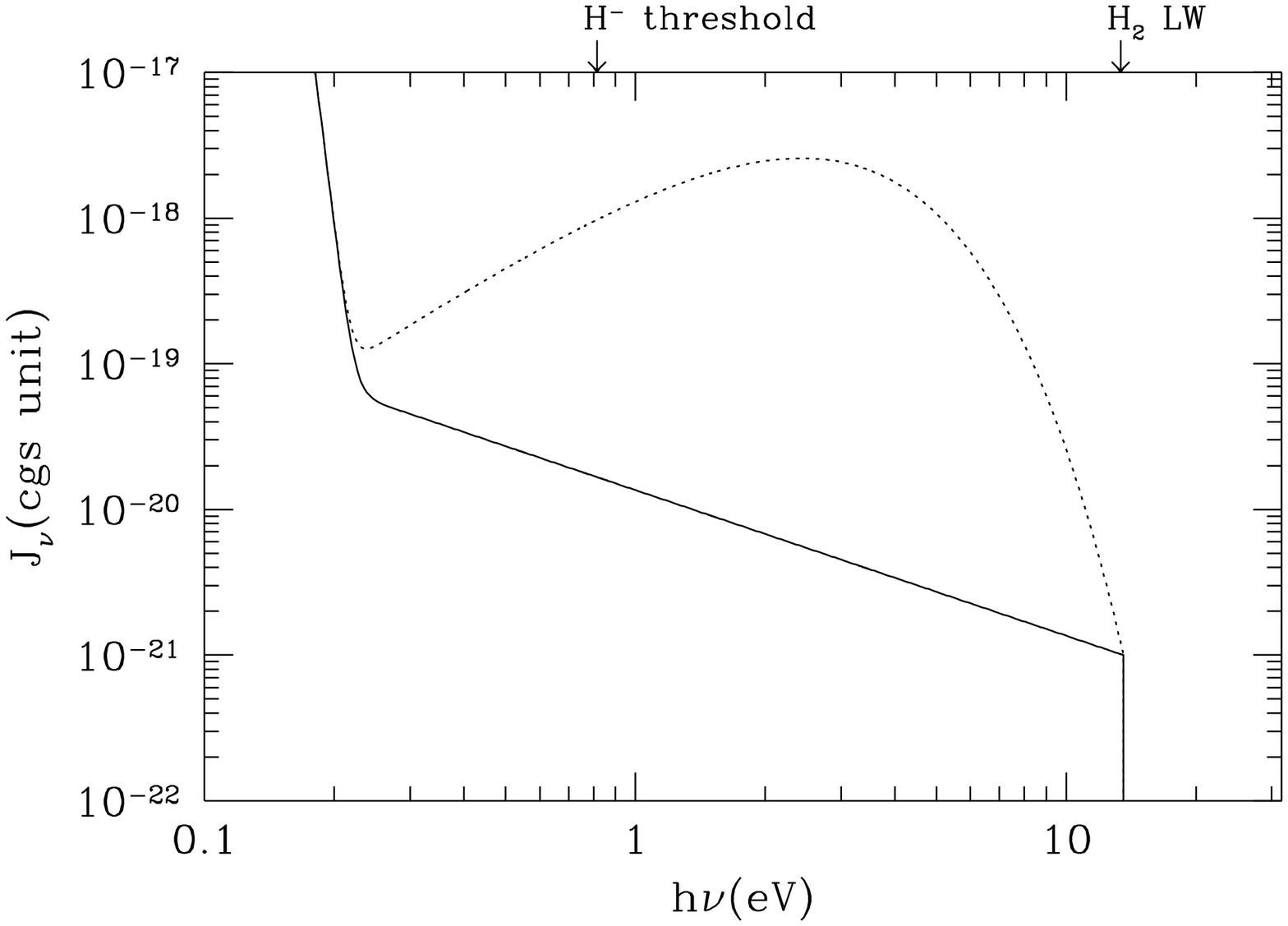]{Two FUV spectra studied in this work. 
The solid and dotted lines indicate those of Types a and b,
respectively.
The normalization of intensity $J_{21}=1$ for both types. 
Also shown are the threshold frequency of H$^{-}$ photodissociation
(0.755 eV) and a representative value of Lyman and Werner (LW) bands
(12.4 eV).  
\label{fig1}}

\figcaption[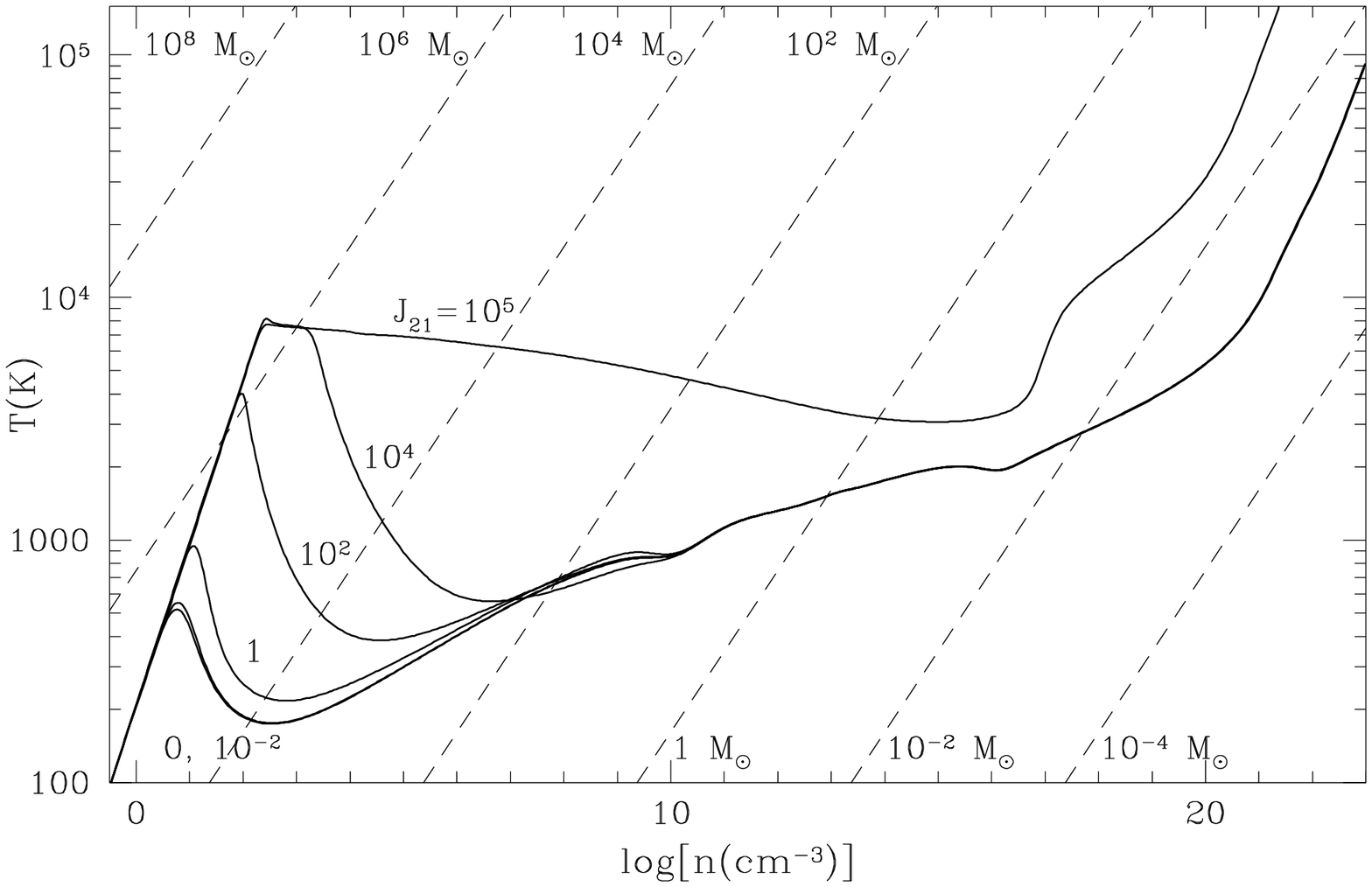]{The density-temperature relation for the collapse of
primordial clouds irradiated with FUV radiation.
The spectra are type {\it a}: $J_{\rm UV}(\nu)=J_{21} \times
10^{-21} (\nu/\nu_{\rm th})^{-1}$. 
The evolutionary trajectories are drawn for $J_{21}=0, 10^{-2}, 1,
10^{2}, 10^{4}$, and $10^{5}$. 
The trajectories for $J_{21}>10^{5}$ are identical to that for
$J_{21}=10^{5}$.
The dashed lines indicate the constant Jeans mass. 
The dark matter gravity is neglected in calculating the Jeans masses.
\label{fig2}}

\figcaption[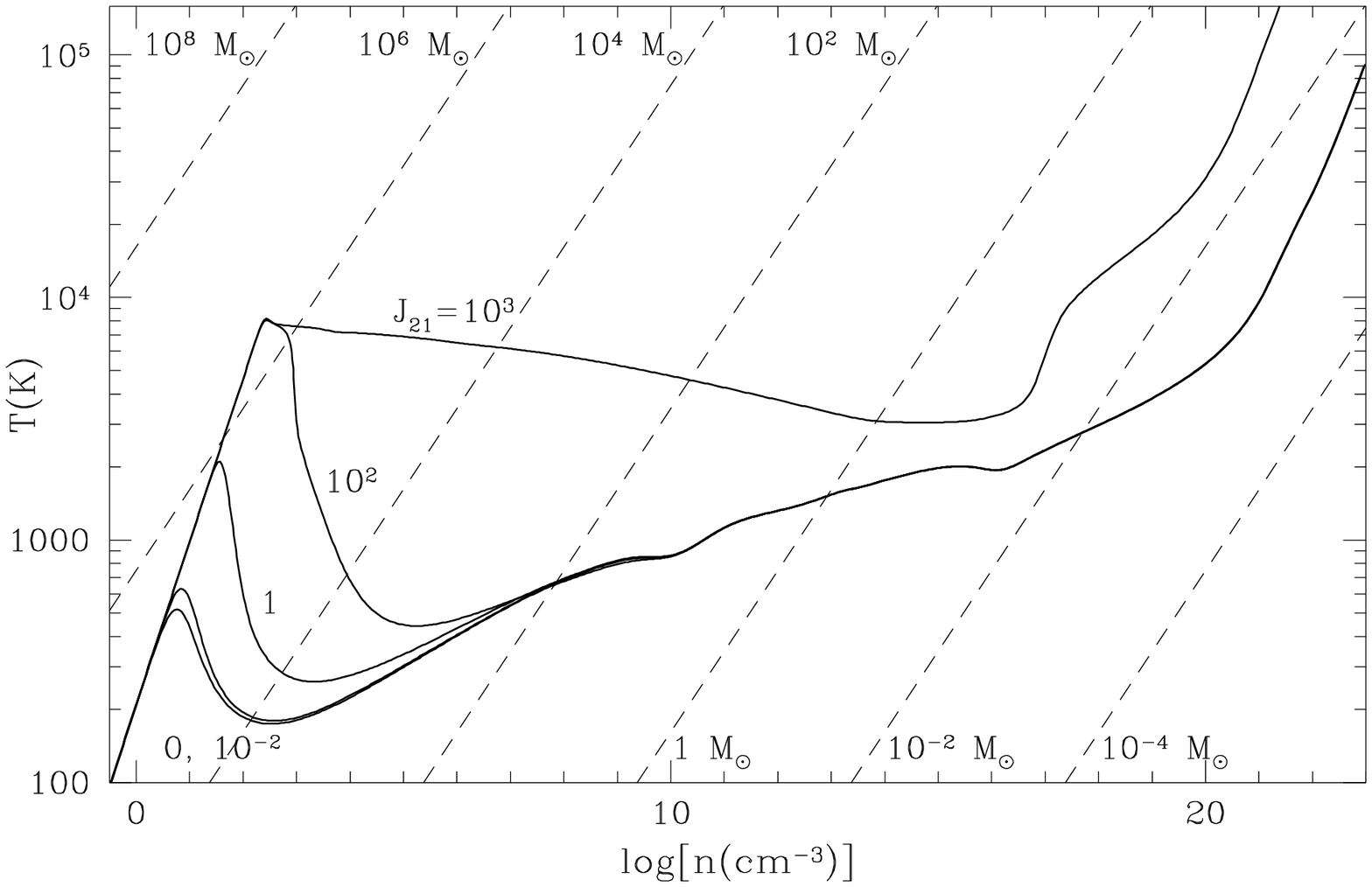]{Same as Fig. 2, but for spectra  
type {\it b}: $J_{\rm UV}(\nu)=J_{21} \times 10^{-21} [B(\nu ;10^{4} {\rm
K})/ B(\nu_{\rm th};10^{4} {\rm K})]  ~~~(\nu < \nu_{\rm th})$.
The evolutionary trajectories are drawn for $J_{21}=0, 10^{-2}, 1,
10^{2}$, and $10^{3}$. 
The trajectories for $J_{21}>10^{3}$ are identical to that for
$J_{21}=10^{3}$.
\label{fig3}}
 
\figcaption[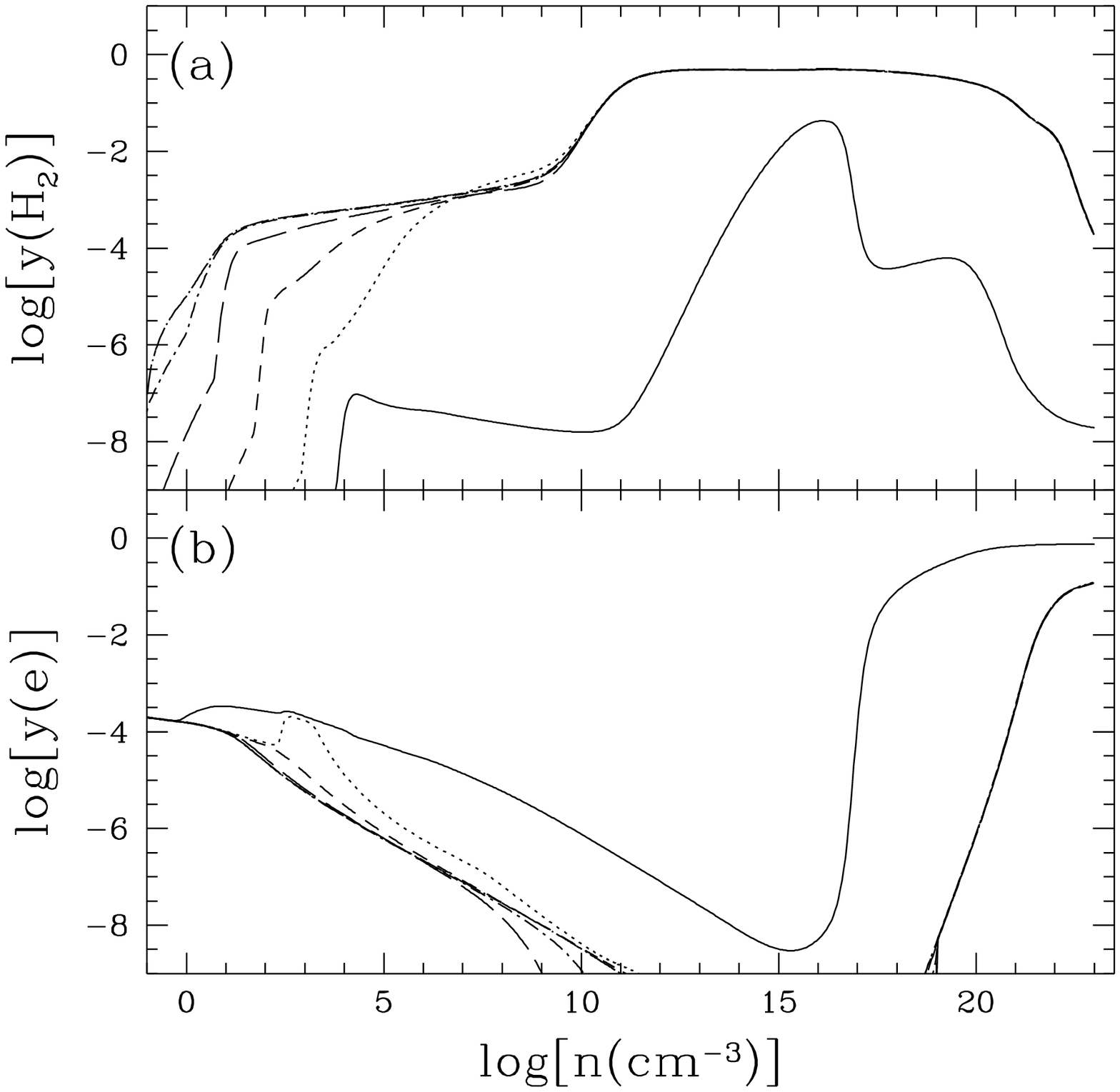]{Concentration of (a) hydrogen molecules $y({\rm
H_2})$ and (b) electrons $y(e)$ for the
clouds irradiated with the power-law type radiation (type {\it a} spectrum in
the text) with $J_{21}=10^{5}$ ({\it solid lines}), $10^{4}$ 
({\it dotted lines}), $10^{2}$({\it short-dashed lines}), 
1 ({\it long-dashed lines}), $10^{-2}$ ({\it dash-dotted lines}), 
and 0 ({\it long-dash-dotted lines}).   
\label{fig4}}

\figcaption[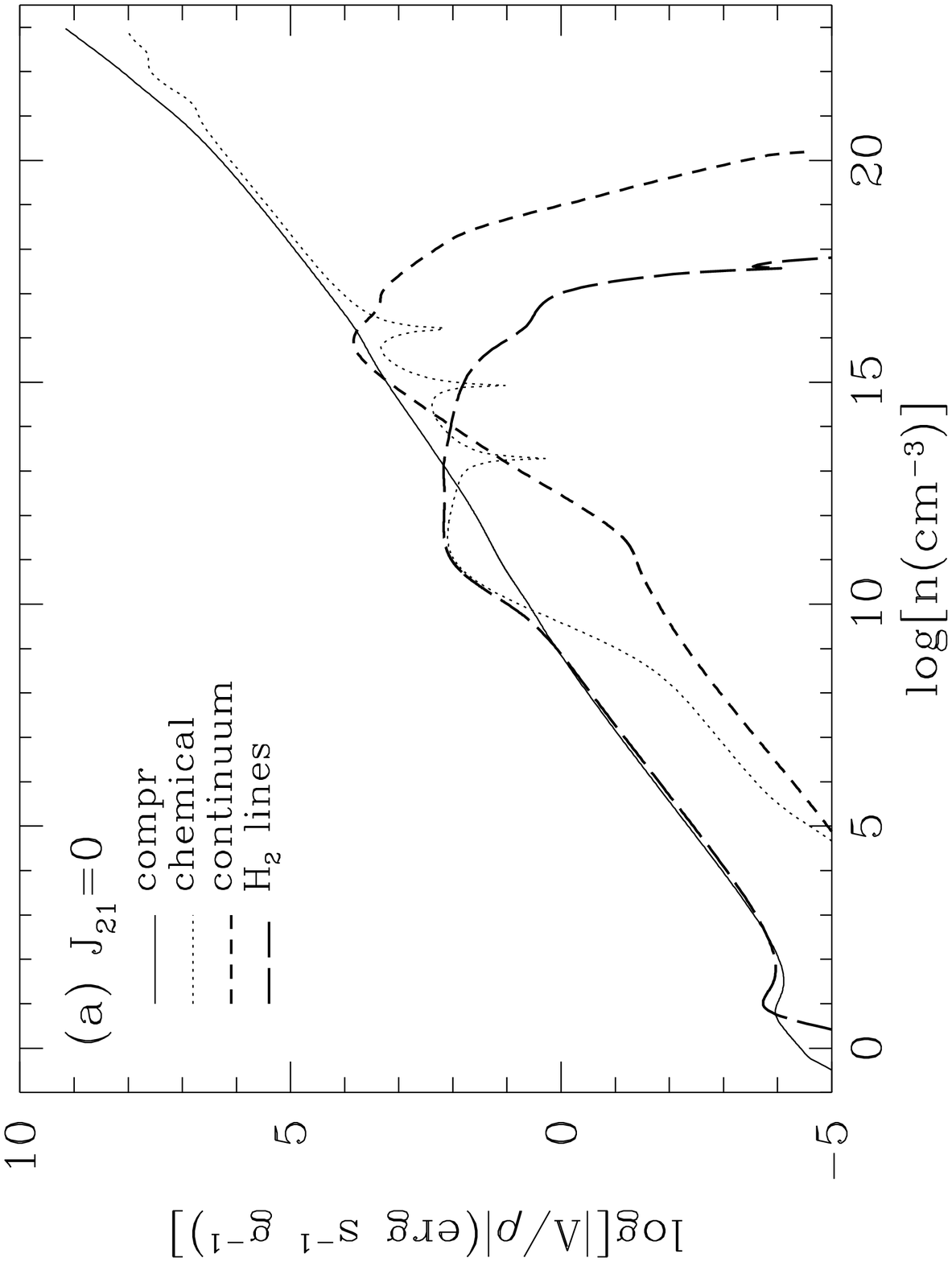,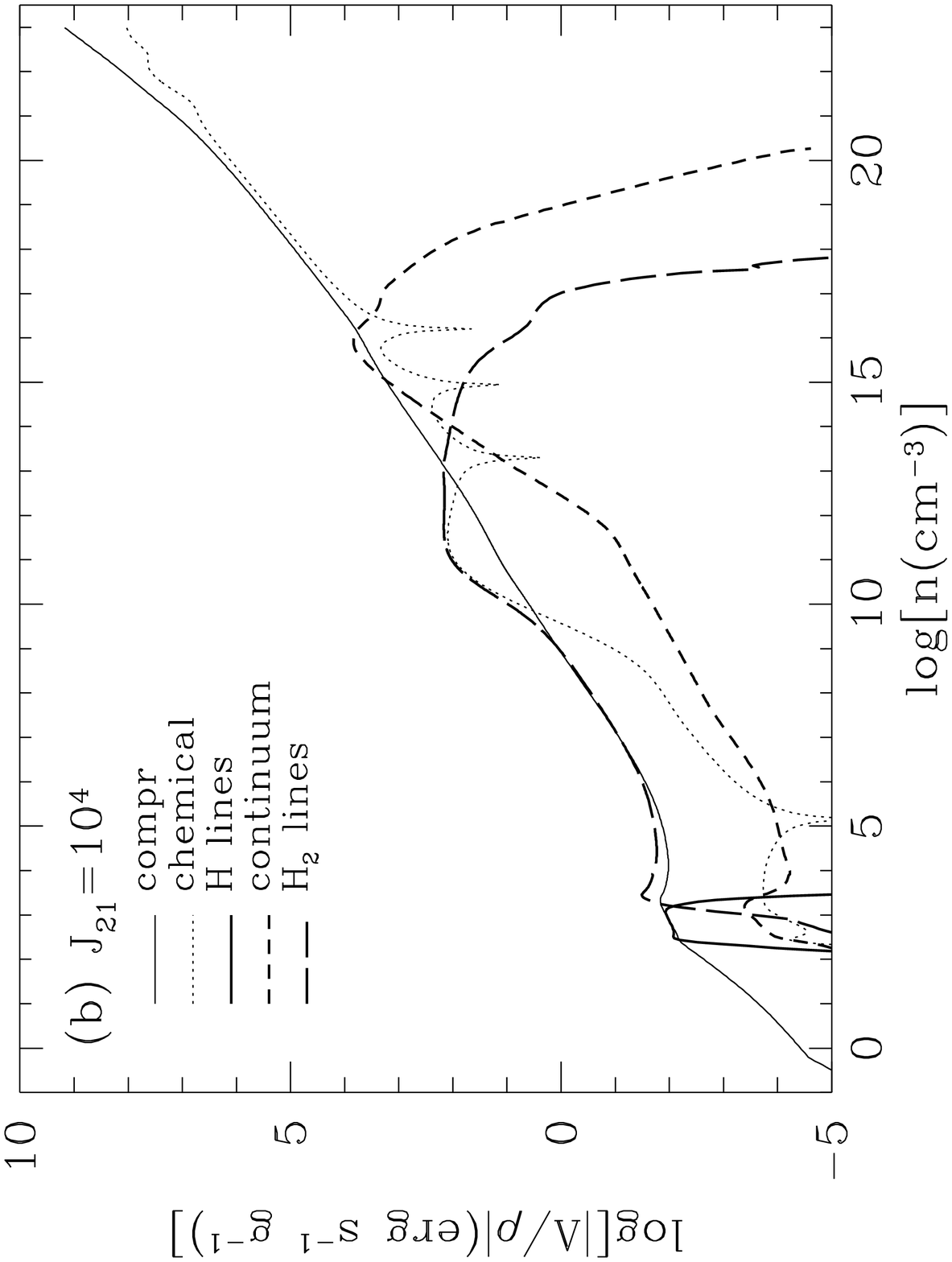,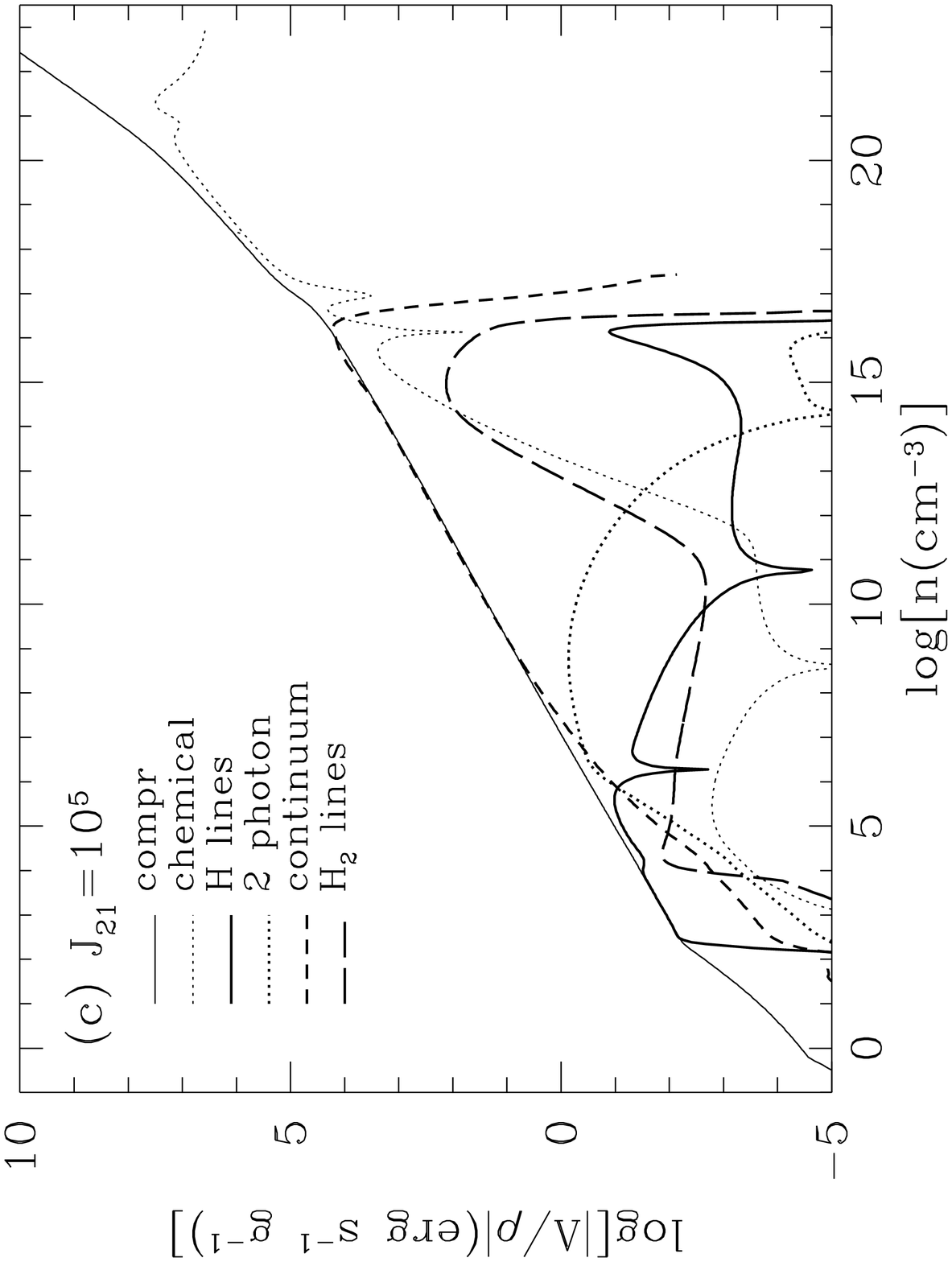]{Cooling/heating rates
per unit baryonic mass as a function of the central number density
for the clouds irradiated with the power-law type radiation of
 (a) $J_{21}=0$, (b) $10^{4}$ and (c) $10^{5}$.  
These include the contributions by the compression, chemical reactions,
atomic hydrogen lines, H two-photon emission, continuum, and
H$_2$ lines. 
The dominant continuum processes are 
(a) H$^{-}$ free-bound emission ($n \la 10^{11.5} {\rm cm^{-3}}$), 
    H$_2$ collision-induced emission ($10^{11.5} {\rm cm^{-3}} \la n$);
(b) H$^{-}$ free-bound emission ($n \la 10^{11.5} {\rm cm^{-3}}$), 
    H$_2$ collision-induced emission ($10^{11.5} {\rm cm^{-3}} \la n$); 
and
(c) H$^{-}$ free-free emission ($10^{2} \la n \la 10^{4.5} {\rm
cm^{-3}}$), H$^{-}$ free-bound emission ($10^{4.5} {\rm cm^{-3}} \la n $),
respectively.
The chemical reactions contributing dominantly to the cooling/heating
rates are:
 (a) and (b): H$_2$ formation ($n \la 10^{13} {\rm cm^{-3}}$),
	 H$_2$ dissociation ($10^{13} \la n \la 10^{15} {\rm
    	cm^{-3}}$), 
	H$_2$ formation ($10^{15} \la n \la 10^{16} {\rm
    	cm^{-3}}$), 
	H$_2$ dissociation ($10^{16} \la n \la 10^{21} {\rm
 	cm^{-3}}$),
   H ionization ($10^{21} {\rm cm^{-3}} \la n$), 
 and
 (c) H$_2$ formation ($n \la 10^{15.5} {\rm cm^{-3}}$),
     H$_2$ dissociation ($10^{15.5} \la n \la 10^{16.5} {\rm
    cm^{-3}}$), 
     H ionization ($10^{16.5} {\rm cm^{-3}} \la n$),
respectively (cf. Fig.4).
 \label{fig5}}

\figcaption[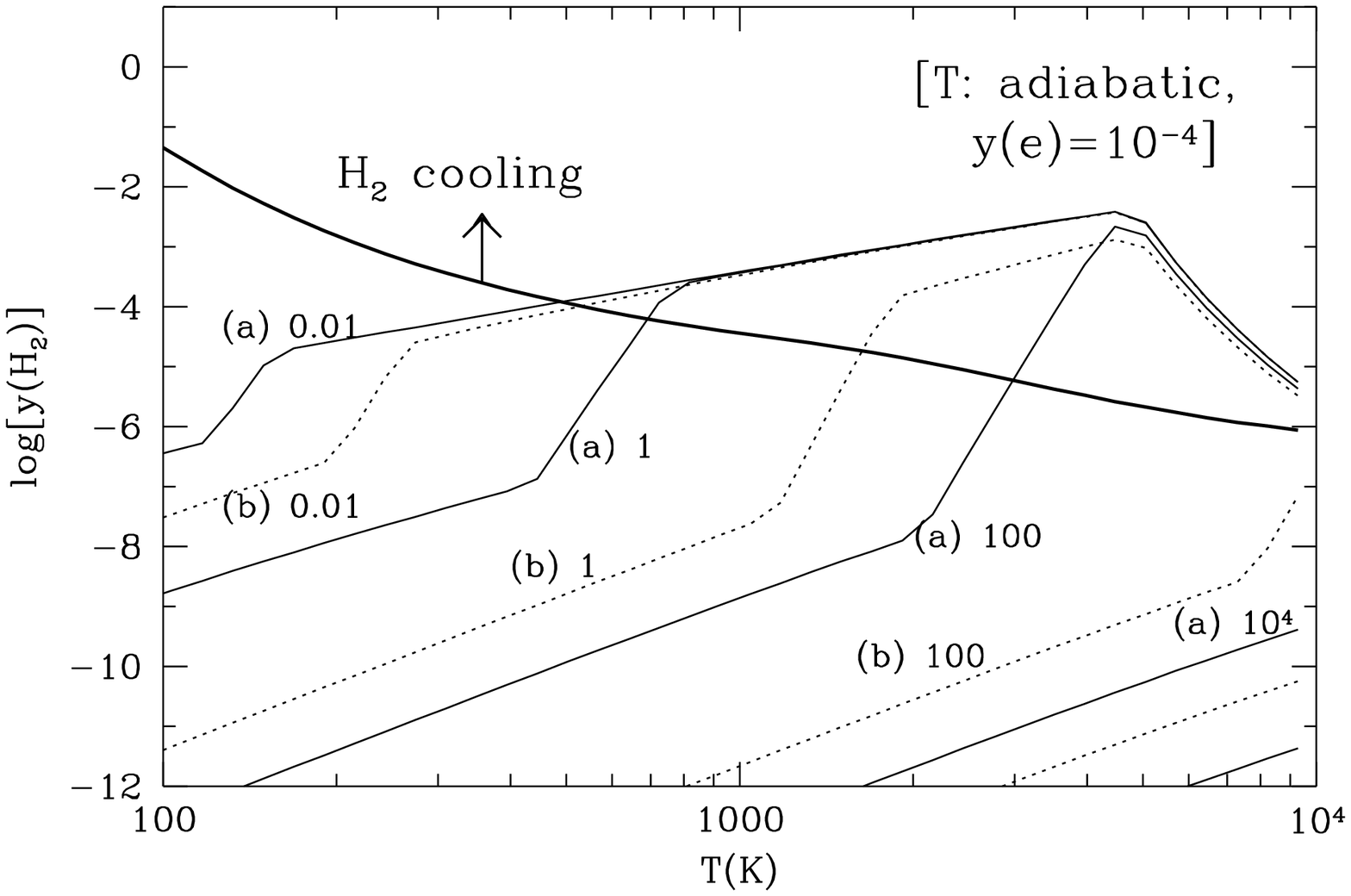]{Estimated H$_2$ fraction vs. the amount of H$_2$ 
needed for efficient H$_2$ cooling during the initial adiabatic
temperature rise. 
We assume that the temperature rises adiabatically from the initial state of
calculation, that is, $T=T_{0}(n/n_{0})^{2/3}$, where $T_{0}=39$ K, and
$n_{0}=8.9 \times 10^{-2} {\rm cm^{-3}}$ are the initial temperature and
number density, respectively.
The ionization degree is taken to be $10^{-4}$.
The H$_2$ fractions are estimated by equation (\ref{eq:yH2}).
The solid and dotted lines illustrate the H$_2$ fractions for Types {\it a}
and {\it b} FUV radiation whose values of $J_{21}$ are indicated in the
figure. 
The thick solid curve shows the necessary H$_2$ fraction for the cloud
 to cool in a free-fall time (eq. [\ref{eq:y_need}])
\label{fig6}
}

\figcaption[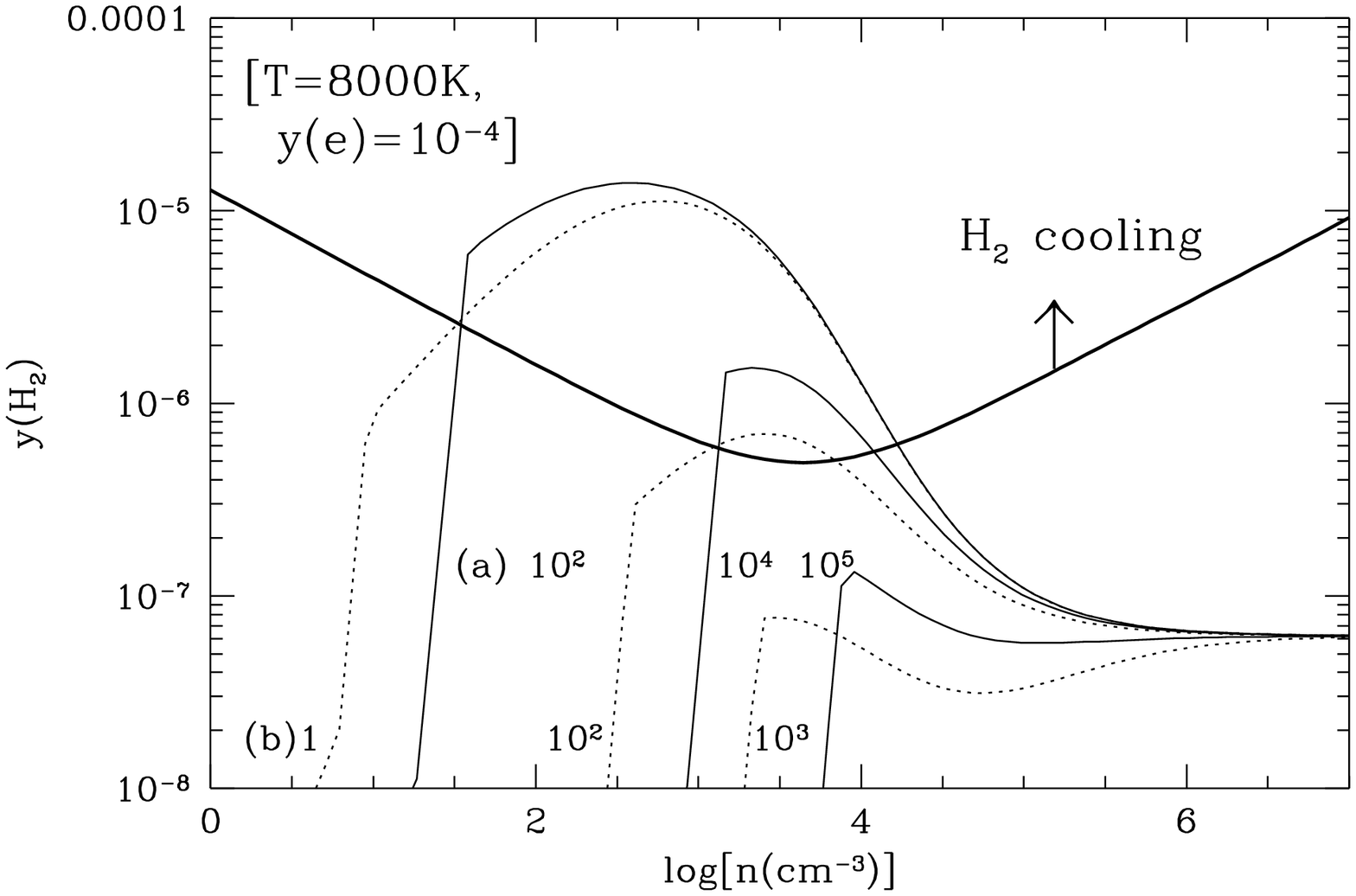]{Estimated H$_2$ fraction vs. the amount of H$_2$ 
needed for efficient H$_2$ cooling for clouds that collapse
isothermally by atomic cooling.
The temperature and ionization degree are taken to be 8000K and $10^{-4}$,
respectively.
The H$_2$ fractions are estimated by eq.(\ref{eq:yH2}).
The solid and dotted lines illustrate the H$_2$ fractions for types {\it a}
and {\it b} FUV radiation whose values of $J_{21}$ are indicated in the
figure. 
The thick solid curve shows the necessary H$_2$ fraction for the cloud
 to cool in a free-fall time (eq.[\ref{eq:y_need}]). 
\label{fig7} }
\end{document}